\documentstyle[12pt]{article}
\if@twoside  m
\else
\fi
 \newcommand{\ie}{{\em i.e.}}
 \newcommand{\eg}{{\em e.g.}}
 \newcommand{\cf}{{\em cf. }}
 
 \newcommand{\etc}{{\em etc. }}
 \newcommand{\Iff}{{\em iff $\:$}}

 \newcommand{\rhs}{{\em rhs }}
 \newcommand{\lhs}{{\em lhs }}
 
 \newcommand{\re}{{\rm Re\,}}
 \newcommand{\im}{{\rm Im\,}}
  \newcommand{\sgn}{{\rm sgn\,}}
 \newcommand{\err}{{\rm Err}}
 \newcommand{\esa}{{\em e.s.a.}}

 \newcommand{\C}{C\!\!\!\rule[.5pt]{.7pt}{6.5pt}\:\:}
 \newcommand{\R}{I\!\!R}
 \newcommand{\Z}{Z\!\!\!Z}
 \newcommand{\BB}{{\cal B}}
 \newcommand{\EE}{{\cal E}}
 \newcommand{\LL}{{\cal L}}
 \newcommand{\OO}{{\cal O}}
 \newcommand{\eps}{\varepsilon}

\begin{document}
\title{Point interactions in a strip}
\date{}
\author{P.~Exner$^{1,2}$,
 R.~Gawlista$^3$,
P.~\v Seba,$^{1,2}$ and M.~Tater$^1$}
\maketitle
\begin{quote}
{\small \em 1 Nuclear Physics Institute, Academy of Sciences, 25068
\v Re\v z near Prague, Czech Republic \\
2 Doppler Institute, Czech Technical University, B\v rehov{\'a} 7,
11519 Prague, Czech Republic
 \\ 3 Lehrstuhl Theoretische Physik I, Fakult\"at f\"ur Physik,
 Ruhr--Universit\"at \\ Bochum, 44780 Bochum--Querenburg, Germany}
\end{quote}

\begin{abstract}
\noindent
We study the behavior of a quantum particle confined to a hard--wall
strip of a constant width in which there is a finite number $\,N\,$
of point perturbations. Constructing the resolvent of the
corresponding Hamiltonian by means of Krein's formula, we analyze its
spectral and scattering properties. The bound state--problem is
analogous to that of point interactions in the plane: since a
two--dimensional point interaction is never repulsive, there are
$\,m\,$ discrete eigenvalues, $\,1\le m\le N\,$, the lowest of which
is nondegenerate. On the other hand, due to the presence of the
boundary the point interactions give rise to infinite series of
resonances; if the coupling is weak they approach the thresholds of
higher transverse modes. We derive also spectral and scattering
properties for point perturbations in several related models: a
cylindrical surface, both of a finite and infinite heigth, threaded
by a magnetic flux, and a straight strip which supports a potential
independent of the transverse coordinate. As for strips with an
infinite number of point perturbations, we restrict ourselves to the
situation when the latter are arranged periodically; we show that in
distinction to the case of a point--perturbation array in the
plane, the spectrum may exhibit any finite number of gaps. Finally,
we study numerically conductance fluctuations in case of random point
perturbations.
\end{abstract}

\tableofcontents

\section{Introduction}

Point interactions were introduced in early days of quantum mechanics
by R.~Kronig, W.G.~Penney, E.~Fermi and others. It lasted several
decades, however, before it became clear how they can be treated in a
mathematically sound way and in what sense they correspond to the
intuitive notion of a $\,\delta$--shaped potential. This in turn made
it possible to develop this method into a powerful tool for analyzing
quantum systems in which the interaction is supported by a family of
sets which are small on an appropriate scale and well separated.

A thorough and extensive discussion of the point interaction method
can be found in the monograph \cite{AGHH}; references to some recent
work are given, \eg, in \cite{AESS,BEKS}. This does not exhaust,
however, the list of possible applications. In this paper we are
going to study another system with point interactions, namely a
quantum particle living in a strip of a fixed width $\,d\,$ with the
Dirichlet (\ie, hard wall) boundary, in which there is a finite
number of pointlike perturbations.

This problem has an obvious physical motivation. The strip is a
natural model for quantum wires which are an object of intensive
experimental and theoretical interest these days. Though the
presently available technologies make it possible to fabricate
mesoscopic structures of extremely pure semiconductor material in
which the electron motion is predominantly ballistic, a real wire is
always non--ideal, containing impurities in the crystallic lattice
which affect its conductivity properties.

Even if one treats the one--electron problem within a suitable range
of energy in the approximation of a particle with the effective mass
whose motion is confined to a strip, the scattering problem in the
presence of impurities is not easy to solve. The assumption that the
impurities have a point character can turn the solution into a more
manageable task; at the same time it is a reasonable approximation
because a typical impurity consists of an alien atom in the lattice
while the wire cross section includes at least several but mostly
many atomic layers. Moreover, an experimental way to produce
{\em artificial} impurities has been reported \cite{KSF}; though the
latter are not exactly pointlike, their existence stresses the need
for better understanding to scattering in quantum waveguides.

Importance of point interactions as models of impurity scattering
in a two--dimensional electron gas was realized long time ago
\cite{Pr}, however, it seems to attract a new attention recently when
several studies of point perturbations on strips, cylinders or tori
has appeared --- see, \eg, \cite{Ba,CBC,KNR} and references
therein. In particular, P.~Bagwell has pointed out the role played by
quasibound states at excited transverse modes which are unstable due
to intermode coupling. Let us remark that similar resonances appear
generically if there is a locally attractive interaction in the wire
(the latter may be of a various origin; it can be induced also in a
purely geometrical way --- see, \eg, \cite{BZ,DES,GJ,VOK,WS}). The
most remarkable manifestation of the resonances are local
deformations of the ideal steplike conductance pattern.

On the other hand, the point scatterers in the above mentioned papers
are treated in a simplified way, usually with the model space
restricted to a finite number of transverse modes. This does not
allow us to introduce the interaction in a proper {\em local} way. In
fact, the authors of Ref.\cite{CBC} discuss the difference between
the situation when the scatterers are attractive or repulsive. It is
known, however, that --- in distinction to one and three dimensions
--- a two--dimensional point interaction obtained as a limit of a
family of scaled potentials is never repulsive; this fact follows
from the logarithmic character of singularity of the corresponding
resolvent kernel --- \cf \cite{AGHH2,GHM} or \cite[Sec.I.5]{AGHH}.

This gives a motivation to analyze the problem of point
perturbations in a strip more carefully; we follow the ideas which
have yielded the one--center scattering matrix discussed in
\cite{ES1}. As in the case without the boundaries \cite{AGHH}, the
point perturbations are constructed as self--adjoint extensions of a
symmetric operator obtained by restriction of the free Hamiltonian in
the strip. Our main tool is the Krein's formula which allows us to
write down the resolvent of the Hamiltonian; the difference from the
free resolvent is a rank $\,N\,$ operator, where $\,N\,$ is the number
of impurities. Since we are interested in local perturbations only,
the number of parameters is again $\,N\;$; they are naturally
interpreted as the corresponding coupling constants.

The spectral problem is then reduced to finding eigenvalues of an
$\,N\times N\,$ matrix. In distinction to the case of point
interactions in a plane, however, the matrix elements are not
elementary functions but sums of certain series. Nevertheless, one
can say a lot about the spectrum. In particular, the mentioned fact
about attractivity of two--dimensional point perturbations is not
affected by the presence of the strip boundary. One can also derive
the S--matrix which yields conductance modulations due to the
impurities, as well as infinite series of resonances associated with
each point perturbation. Another feature observed in scattering are
turbulent patterns of the probability flow caused by the impurities,
which have been noticed in Ref.\cite{CBC} and in \cite{BZ,WS} in a
bent--guide context.

Furthermore, strips with point perturbations offer a useful test field
for some properties dependent on the dimension of the configuration
space, such as number of open gaps in a periodic situation, or
existence of localization in case of random perturbations, because,
roughly speaking, they are ``halfway'' between the one--dimensional
and two--dimensional situation. We shall show that while a straight
polymer in the plane has at most one gap \cite[Sec.III.4]{AGHH}, a
``coated polymer'', \ie, a periodic array of point perturbations in a
strip can exhibit any finite number of open gaps. This raises the
question, of course, whether the Bethe--Sommerfeld conjecture can be
proven in this situation.

Finally, if the perturbation positions or coupling constants are
distributed at random, we can study statistical properties of
observable quantities such as conductance of a wire with point
perturbations. The explicit solvability of our model makes it
suitable for comparison with predictions of random--matrix theory;
this will be done in the concluding section.

\section{A one--center perturbation in an infinite strip}

Consider a straight planar strip $\,\Omega:= \R\times [0,d]\,$ and
suppose that the free motion in $\,\Omega\,$ is governed by the
Dirichlet Laplacian \cite[Sec.XIII.15]{RS}; since the boundary of
$\,\Omega\,$ is smooth, it acts as $\,H_0\psi= -\Delta\psi\,$ on the
domain of all $\,\psi\,$ for which the \rhs (in the sense of
distributions) belongs to $\,L^2(\Omega)\,$ and which satisfy the
boundary conditions
   \begin{equation} \label{strip boundary}
\psi(x,0)\,=\, \psi(x,d)\,=\,0
   \end{equation}
for all $\,x\in\R\,$. Unless stated otherwise, we put for simplicity
$\,d=\pi\;$; the results for a general strip width are then obtained
by a simple scaling transformation.

\subsection{Boundary conditions}

Our first goal is to construct a one--center singular perturbation to
$\,H_0\,$ supported by a point $\,\vec a:=(a,b)\,$ with $\,b\in
(0,\pi)\,$. This can be done using the procedure mentioned
in the introduction; since locally  there is no difference between
point interactions in the plane and in the strip, we can use the
standard boundary conditions \cite[Chap.I.5]{AGHH}. We introduce the
regularized boundary values
$$
L_0(\psi,\vec a)\,:=\, \lim_{|\vec x-\vec a|\to 0}\, {\psi(\vec
x)\over \ln |\vec x\!-\!\vec a|}\,, \quad\; L_1(\psi,\vec a)\,:=\,
\lim_{|\vec x-\vec a|\to 0} \bigl\lbrack \psi(\vec x)- L_0(\psi,\vec
a)\, \ln |\vec x\!-\!\vec a| \bigr\rbrack
$$
and require
   \begin{equation} \label{bc}
L_1(\psi,\vec a)\,+\, 2\pi\alpha L_0(\psi,\vec a)\,=\,0\,.
   \end{equation}
To any $\,\alpha\in\R\,$ we define the self--adjoint operator
$\,H(\alpha, \vec a)\,$ as
   \begin{eqnarray} \label{one-center perturbation}
H(\alpha,\vec a)\psi \!\!&:=&\!\! -\Delta\psi\,, \nonumber \\ \\
D(H(\alpha,\vec a)) \!\!&:=&\!\! \Bigl\lbrace\, \psi\,:\:
-\Delta\psi \;\;{\rm is\;}\; L^2\:\; {\rm on}\;\;
\Omega\!\setminus\!\{\vec a\}\;\; {\rm and\; (\ref{strip
boundary}), (\ref{bc})\; are\; satisfied\,} \Bigr\rbrace\,; \nonumber
   \end{eqnarray}
here again $\,-\Delta\psi\,$ is understood in the sense of
distributions. The case $\,\alpha=\infty\,$, \ie $\;L_0(\psi,\vec
a)=0\,$, corresponds to the free Hamiltonian $\,H_0\,$.

\subsection{The resolvent}

To find the resolvent of $\,H(\alpha,\vec a)\,$ we start with that of
$\,H_0\,$. We use the decomposition into transverse modes,
$\,L^2(\Omega)\,=\, \bigoplus_{n=1}^{\infty}\, L^2(\R)\otimes
\{\chi_n\}\,$, where $\,\chi_n(y):= \sqrt{2\over\pi}\, \sin(ny)\;$;
then the free Hamiltonian can be written as
   \begin{equation} \label{Hamiltonian decomposition}
H_0\,=\, \bigoplus_{n=1}^{\infty}\, h_n\otimes I_n\,, \quad\; h_n\,:=\,
-\,{d^2\over dx^2}\,+\,n^2
   \end{equation}
with $\,D(h_n):= AC^2(\R)\,$. It follows that the free resolvent is
an integral operator with the kernel
   \begin{equation} \label{free kernel}
G_0(\vec x_1,\vec x_2;z)\,\equiv\, (H_0\!-z)^{-1}(\vec x_1,\vec
x_2)\,=\, {i\over\pi}\, \sum_{n=1}^{\infty}\,
\frac{e^{ik_n(z)|x_1-x_2|}} {k_n(z)}\, \sin(ny_1)\, \sin(ny_2)\,,
   \end{equation}
where $\,\vec x_j=(x_j,y_j)\,$ and $\,k_n(z):=\, \sqrt{z\!-\!n^2}\,$.
The complex energy here belongs to the resolvent set of $\,H_0\,$,
$\;z\in \C\setminus[1,\infty)\;$; the function
$\,G_0(\cdot,\cdot;z)\,$ is defined and smooth everywhere outside the
hyperplane $\,\vec x_1= \vec x_2\,$, but the sum on the \rhs may not
converge absolutely if the longitudinal coordinates coincide,
$\,x_1=x_2\,$. Moreover, the \rhs makes sense also for all
non--integer $\,z>1\,$, where it gives the boundary value of the
kernel at the cut; one has to choose properly the branch of the
square root in $\,k_n(z)\,$. We note for further purposes that
for real $\,z\,$ below the lowest threshold, $\,z<1\,$, the kernel is
strictly positive, $\,G_0(\vec x_1,\vec x_2;z) >0\,$ for all mutually
different $\,\vec x_1,\vec x_2 \in\Omega\,$ --- \cf \cite[App. to
Sec.XIII.12]{RS}.

The sought kernel of the full resolvent can be obtained by means of
the Krein formula \cite[App.A]{AGHH}, which yields the Ansatz
   \begin{equation} \label{Krein}
(H(\alpha,\vec a)\!-\!z)^{-1}(\vec x_1,\vec x_2)\,=\, G_0(\vec
x_1,\vec x_2;z) +\lambda(\alpha,\vec a;z) G_0(\vec x_1,\vec a;z)
G_0(\vec a,\vec x_2;z)\,.
   \end{equation}
To determine the function $\,\lambda\,$ we use the fact
that, by definition, $\,(H(\alpha,\vec a)\!-\!z)^{-1}\,$
maps into the domain of $\,H(\alpha,\vec a)\,$. Writing $\,\psi:=
(H(\alpha,\vec a)\!-\!z)^{-1}\varphi\,$, $\,\psi_0:=
(H_0\!-\!z)^{-1}\varphi\,$ for an arbitrary $\,\varphi\in
L^2(\Omega)\,$, we get from the last relation
   \begin{equation} \label{decomposition}
\psi(x,y)\,=\, \psi_0(x,y)\,+\, {i\lambda\over\pi}\,
\sum_{n=1}^{\infty}\, \frac{e^{ik_n(z)|x-a|}} {k_n(z)}\, \sin(ny)\,
\sin(nb)\, \psi_0(a,b)\,.
   \end{equation}
Since $\,\psi_0\in D(H_0)\,$ is smooth at $\,\vec x=\vec a\,$, the
generalized boundary values can be written as $\,L_j(\psi,\vec a)=
\LL_j(\vec a)\psi_0(\vec a)\,$. The singularity at the ``diagonal''
of the resolvent kernel for planar regions with a smooth boundary is
well known \cite[Chaps.11 and 14]{Ti}. It is not difficult, however,
to find it directly: we have
   \begin{eqnarray*}
\lefteqn{\LL_0(\vec a)\,=\,\lim_{|\vec x-\vec a|\to 0}\, {1 \over \ln
|\vec x\!-\!\vec a|}\, \left\lbrace\,1\,+\, {i\lambda\over\pi}\,
\sum_{n=1}^{\infty}\, \frac{e^{ik_n|x-a|}} {k_n}\, \sin(ny)\,
\sin(nb) \,\right\rbrace} \\ \\ &&
=\, {i\lambda\over\pi}\, \lim_{u\to 0}\,{1\over\ln u}\,
\sum_{n=1}^{\infty}\, \frac{e^{ik_nu}}{k_n}\, \sin^2(nb) \,=\,
{\lambda\over\pi}\, \lim_{u\to 0}\,{1\over\ln u}\,
\sum_{n=1}^{\infty}\, \frac{e^{-(n-z/2n)u}}{n}\, \sin^2(nb)\,,
   \end{eqnarray*}
because the differences $\,k_n^{-1}\!-(in)^{-1}\,$ and
$\,e^{ik_n+n-z/2n}\!-1\,$ are $\,\OO(n^{-3})\,$. The limit is easily
computed giving $\,\LL_0(\vec a)\,=\, -\lambda/2\pi\;$; this in turn
yields
$$
\LL_1(\vec a)\,=\, 1\,+\,{i\lambda\over\pi}\, \sum_{n=1}^{\infty}\,
\left(\, {\sin^2(nb)\over k_n}\,-\, {1\over 2in} \,\right)\,.
$$
Using now the boundary conditions (\ref{bc}) we find that
   \begin{equation} \label{lambda}
\lambda(\alpha,\vec a;z)\,=\, \Lambda(\alpha,\vec a;z)^{-1}
   \end{equation}
with
   \begin{equation} \label{Lambda}
\Lambda(\alpha,\vec a;z)\,=\, \alpha\,+\,{1\over\pi i}\,
\sum_{n=1}^{\infty}\, \left(\, {\sin^2(nb)\over k_n(z)}\,-\, {1\over
2in} \,\right)\,=:\, \alpha\,-\, \xi(\vec a,z)\;;
   \end{equation}
for the sake of brevity we shall sometimes drop a part or all of the
arguments in the following.

\subsection{Scaling behavior}

Although the modification of these results for a general strip is
simple, it is worthy of a brief comment. The logarithmic factor in the
definition of the regularized boundary values refers to a fixed
length scale. If we choose the natural scale of our problem for it,
we replace the factor by $\,\ln|\nu(\vec x-\vec a)|\,$, where
$\,\nu:= \pi/d\,$. In that case $\,\LL_0(\vec a)\,$ does not
change, while $\,\LL_1(\vec a)\,$ acquires in addition to the scaling
also the additive factor $\,{1\over 2\pi}\, \ln\left(\pi\over
d\right)\,$. The formula (\ref{lambda}) reads then $\,\lambda(
\alpha,\vec a;z)= (\alpha- \xi_d(\vec a,z))^{-1}\,$ with
   \begin{equation}  \label{scaling}
\xi_d(\vec a,z)\,=\, {i\over d}\, \sum_{n=1}^{\infty}\, \left(\,
{\sin^2(nb\nu)\over k_n(z)}\,-\, {d\over 2\pi in}
\,\right)\,+\, {1\over 2\pi}\, \ln\left(\pi\over d\right)\,,
   \end{equation}
where $\,k_n(z)\,:=\,\sqrt{z-(\pi n/d)^2}$. Hence changing the strip
width is equivalent to a shift in the coupling constant for a fixed
$\,d\;$; one has to replace $\,\alpha\,$ by $\,\alpha\,-\, {1\over
2\pi}\, \ln\left(\pi\over d\right)\,$. In particular, making the
strip thin with $\,\alpha\,$ preserved corresponds to choosing a
large positive $\,\alpha\,$ in the rescaled problem.

\subsection{The resolvent \`a la Zorbas}

The resolvent kernel can be obtained even without the boundary
conditions (\ref{bc}) if we employ an integral formula due to Zorbas
\cite[Thm.4.1]{Zo}. According to it, the operator family in question
can be parametrized by a real $\,\theta\;$; the function
$\,\Lambda(\alpha,\vec a;z)\,$ is replaced by
   \begin{eqnarray*}
\lefteqn{\Lambda(\theta,\vec a;z)\,=\,
(1\!+\!e^{i\theta})^{-1}\,\biggl\lbrace (i\!-\!z)\, \int_{\Omega} G_0(\vec
x,\vec a;z) G_0(\vec x,\vec a;i)\, d\vec x} \\ \\ && \phantom{AAAA}
-\, e^{i\theta}(i\!+\!z)\, \int_{\Omega} G_0(\vec x,\vec a;z)
G_0(\vec x,\vec a;-i)\, d\vec x \,\biggr\rbrace\,. \phantom{AAAAAA}
   \end{eqnarray*}
Using the formula (\ref{free kernel}) together with the identities
$\,i\!-\!z= k_n(i)^2\!-k_n(z)^2\,$, $\,i\!+\!z= k_n(z)^2\!
-k_n(-i)^2\;$ the \rhs can be easily computed; we get
   \begin{equation} \label{Zorbas}
\Lambda(\theta,\vec a;z)\,=\, {1\over\pi i}\, \sum_{n=1}^{\infty}\,
\left\lbrace\, {1\over k_n(z)}\,-\, i\im\left({1\over k_n(i)}\right)
\,+\, {i\sin\theta\over 1\!+\!\cos\theta}\, \re\left({1\over
k_n(i)}\right)  \right\rbrace\, \sin^2(nb)\,.
   \end{equation}
Since $\,\re\left(k_n(z)^{-1}\right)\,$ as well as $\,k_n(z)\!
-\!k_n(i)\,$ are $\,\OO(n^{-3})\,$, the series converges and can be
splitted into the $\,\theta$--independent and $\,\theta$--dependent
parts. Comparing the two parametrizations, we find
   \begin{equation} \label{comparison}
\alpha(\theta)\,=\, F(b)\,+\, {\sin\theta\over 1\!+\!\cos\theta}\,
G(b)\,,
   \end{equation}
where
   \begin{eqnarray*}
&& F(b)\,:=\, {1\over\pi}\, \sum_{n=1}^{\infty}\, \left(
\sqrt{{\sqrt{n^4\!+1}+n^2\over 2(n^4\!+1)}}\,\sin^2(nb)\,-\, {1\over
2n} \right)\,, \\ \\ &&
G(b)\,:=\, {1\over\pi}\, \sum_{n=1}^{\infty}\,
\sqrt{{\sqrt{n^4\!+1}-n^2\over 2(n^4\!+1)}}\,\sin^2(nb)\;;
   \end{eqnarray*}
the function $\,\alpha\,$ clearly maps $\,(-\pi,\pi)\,$ bijectively
onto $\,\R\,$.

\subsection{Spectral properties}

As a rank--one perturbation in the resolvent, the point interaction
does not change the essential spectrum, \ie, we have $\,\sigma_{ess}
(H(\alpha,\vec a))= [1,\infty)\,$ with the multiplicity
$\,2[\sqrt{E/n^2}]\,$ at a point $\,E\,$, where $\,[\,\cdot\,]\,$
denotes the integer part.

To determine the point spectrum, one has to find poles of the
resolvent, \ie, to solve the equation
   \begin{equation} \label{pole condition}
\xi(\vec a,z)\,=\,\alpha
   \end{equation}
for $\,z\in\R\,$. First consider the case $\,z<1\,$. It is useful to
introduce $\,\kappa_n(z):= -ik_n(z)= \sqrt{n^2-z}\;$; then the
function $\,\xi(\vec a,\cdot)\,$ can be expressed as
   \begin{eqnarray*}
\xi(\vec a,z) &\!=\!& {1\over\pi}\, \sum_{n=1}^{\infty} \left(
{\sin^2(nb)\over \kappa_n(z)}\,-\,{1\over 2n} \right) \\ \\
&\!=\!& \xi(\vec a,z')\,+\, {1\over\pi}\, \sum_{n=1}^{\infty}\,
{\kappa_n(z')- \kappa_n(z)\over \kappa_n(z)\kappa_n(z')}\,
\sin^2(nb)\,.
   \end{eqnarray*}
The sum at the right hand side converges because the coefficients at
$\,\sin^2(nb)\,$ decay like $\,\OO(n^{-3})\,$ as $\,n\to\infty\,$. In
particular, one can compute easily
   \begin{equation} \label{xi at zero}
\xi(\vec a,0) \,=\, {1\over 2\pi}\, \ln(2\sin b)\;;
   \end{equation}
this quantity is positive close to the center of the strip, $\,b\in
\left( {\pi\over 6},\, {5\pi\over 6} \right)\,$, and nonpositive
otherwise. One has
$$
{\partial\xi \over\partial z}\,=\, {1\over 2\pi}\,
\sum_{n=1}^{\infty}\, {\sin^2(nb)\over \kappa_n(z)^3}\,>\, 0\,,
$$
so the function is monotonously growing in $\,(-\infty,1)\;$; it
diverges at both endpoints. We have
$$
\xi(\vec a,z)\,=\, {\sin^2 b\over \pi\sqrt{1-z}}\,+\,\OO(1) \qquad
{\rm as} \qquad z\to 1-\,,
$$
while on the negative side the definition formula shows that
$\,\xi(\vec a,\cdot)\,$ diverges logarithmically. A natural
conjecture about its asymptotics is
$$
\xi(\vec a,z)\,=\, -\,{1\over 4\pi}\, \ln\left( -\,{z\over 4} \right)
\,+\, {1\over 2\pi}\, \Psi(1)\,+\,\OO(1) \qquad {\rm as} \qquad z\to
-\infty\,,
$$
where $\,\Psi(1)= -\gamma= 0.57721\;$ (compare with
\cite[Sec.I.5]{AGHH}). Figure~1   
   \setcounter{figure}{0}
   \begin{figure} \label{xi}
   \caption{The function $\,\xi\,$ for different values of $\,b\,$.}
   \end{figure}
suggests that it is indeed the case. The asymptotics is independent
of $\,b\,$ but not uniformly in $\,a\,$ as the relation (\ref{xi at
zero}) shows.

Another property of $\,\xi(\vec a,\cdot)\,$ is its monotonicity
across a halfstrip,
$$
\xi(\vec a,z)\,>\, \xi(\vec a',z) \qquad {\rm iff} \qquad \left|
b-\,{\pi\over 2}\right|\,<\, \left| b'-\,{\pi\over 2}\right|\,.
$$
To prove it, we employ the identity $\,\sin^2(nb)-\sin^2(nb')=
\sin(n(b+b')) \sin(n(b-b'))\,$ for $\,0<b'<b\le\, {\pi\over 2}\;$; it
shows that
$$
\xi(\vec a,z)- \xi(\vec a',z)\,=\, G_0(0,b+b';0,b-b';z)\,,
$$
so the result follows from the above mentioned positivity of the
free-resolvent kernel.

Properties of solutions to Eq.(\ref{pole condition}) follow easily
from this discussion: to any coupling constant $\,\alpha\in\R\,$
there is {\em just one eigenvalue} $\;\epsilon(\alpha,\vec a)\,$ in
$\,(-\infty,1]\,$. The function $\,\epsilon(\cdot,\vec a)\,$ is
monotonously growing,
   \begin{equation} \label{coupling monotonicity}
\epsilon(\alpha,\vec a)\,>\, \epsilon(\alpha',\vec a) \qquad {\rm
iff} \qquad \alpha>\alpha'\,,
   \end{equation}
and has the following asymptotic behavior,
   \begin{eqnarray} \label{energy asymptotics}
\epsilon(\alpha,\vec a) &\!=\!& 1\,-\, \left( {\sin^2 b\over
\pi\alpha}\right)^2 \,+\, \OO(\alpha^{-3}) \qquad {\rm as} \qquad
\alpha\to\infty\,, \nonumber \\ \\
\epsilon(\alpha,\vec a) &\!\approx\!& -4\,e^{2(\Psi(1)- 2\pi\alpha)}
\qquad {\rm as} \qquad \alpha\to-\infty\,. \nonumber
   \end{eqnarray}
Hence $\,H(\alpha,\vec a)\,$ preserves the property of the point
interaction in the plane, namely that it has always a bound state
except in the free case; also the asymptotics for
$\,\alpha\to-\infty\,$ is the same. On the other hand, proximity of
the boundary pushes the bound--state energy up,
   \begin{equation} \label{halfwidth monotonicity}
\epsilon(\alpha,\vec a)\,<\, \epsilon(\alpha,\vec a') \qquad {\rm
iff} \qquad \left| b-\,{\pi\over 2}\right|\,<\, \left| b'-\,{\pi\over
2}\right|\,.
   \end{equation}
The resolvent formula (\ref{Krein}) provides us also with the
(non--normalized) wavefunction of the bound state through the residue
at the pole, $\,\psi(\cdot;\alpha,\vec a)= G_0(\cdot,\vec a;
\epsilon(\alpha,\vec a))\,$, so we have
   \begin{equation} \label{bs wavefunction}
\psi(x;\alpha,\vec a)\,=\,
{1\over\pi}\, \sum_{n=1}^{\infty}\,
\frac{e^{-\kappa_n(\epsilon)|x-a|}}
{\kappa_n(\epsilon)}\, \sin(ny)\, \sin(nb)\,.
   \end{equation}
If $\,\,\alpha\,$ decreases, $\,\psi\,$ becomes well localized and
resembles the Hankel eigenfunction of the point interaction in the
plane, while in the limit $\,\alpha\to\infty\,$ we have
$$
\psi(x;\alpha,\vec a)\,\approx\, \alpha\, {\sin y\over\sin b}\,
e^{-|x-a|\sin^2 b/\pi\alpha}\,+\, {1\over\pi}\, \sum_{n=2}^{\infty}\,
\frac{e^{-|x-a|\,\sqrt{n^2-1}}} {\sqrt{n^2-1}}\, \sin(ny)\,
\sin(nb)\;;
$$
the leading term here is the product of $\,\chi_1(y)\,$ with the
eigenfuction of the {\em one--dimensional} attractive point
interaction of the strength $\, -(2/\pi\alpha)\sin^2 b\;$; recall
that in view of the preceding section this corresponds by rescaling
to the limit $\,d\to 0\,$.
   \begin{figure} \label{1EV}
   \caption{Eigenvalue plots for different values of $\,
\alpha,\,b\,$.}
   \end{figure}

   \begin{figure} \label{1EF}
   \caption{Eigenfunctions for different values of $\,b\,$.}
   \end{figure}
These results are illustrated on Figs.~2 and 3.  

There are {\em no eigenvalues embedded in the continuous spectrum.}
The function $\,\xi(\vec a,\cdot)\,$ is defined everywhere in
$\,(1,\infty)\,$ except at the thresholds, $\,z=m^2\,$, but it is now
complex because $\,\kappa_n(z)\,$ is negative imaginary for the open
channels. In particular,
   \begin{equation} \label{im part}
\im \xi(\vec a,z)\,=\, {1\over\pi}\, \sum_{n=1}^{[\sqrt{z}]}\,
{\sin^2(nb)\over k_n(z)}\,>\,0
   \end{equation}
for a non--integer $\,z>1\,$, so the function $\,a-\xi(\vec
a,\cdot)\,$ has no zeros there. The argument does not apply if
$\,z=m^2\,$ because the corresponding $\,k_m(z)=0\,$ and none of the
expressions at the \rhs of (\ref{Krein}) makes sense. However, one
can check directly that the singularities cancel mutually. Around
$\,z=m^2\,$, the free resolvent kernel behaves as
$$
G(\vec x_1,\vec x_2;z)\,=\, {i\over\pi}\, {\sin(my_1)\sin(my_2) \over
\sqrt{z-m^2}}\, \err(z)\,,
$$
where $\,\err(z)\,$ stands as a shorthand for
$\,1+\OO(\sqrt{z-m^2})\,$, and
$$
\Lambda(\alpha,\vec a;z)\,=\, \tilde\alpha\,-\, {i\over\pi}\,
{\sin^2(mb) \over \sqrt{z-m^2}}\,+\, \OO(\sqrt{z-m^2})
$$
with $\,\tilde\alpha:= \alpha+\, {1\over 2\pi m} \,-\, {i\over\pi}\,
\sum_{n\ne m} \left(\, {\sin^2(nb)\over k_n(m^2)}\,-\, {1\over
2in} \,\right)\,$. Hence the full--resolvent kernel behaves
asymptotically as
$$
{i\tilde\alpha\over\pi}\, {\sin(my_1)\sin(my_2) \over
\sqrt{z-m^2}} \,\left( \tilde\alpha\,-\, {i\over\pi}\, {\sin^2(mb)
\over \sqrt{z-m^2}}\,\right)^{-1}\, \err(z)
$$
and cannot have a pole at $\,z=m^2\,$. In conclusion, no real $\,z\ge
1\,$ is an eigenvalue of $\,H(\alpha,\vec a)\,$. On the other hand,
the resolvent has resonance zeros in the complex plane as we shall
show in Section~2.7 below.

\subsection{Scattering}

The existence of the wave operators is easy to establish because the
point interaction is a rank--one perturbation in the resolvent, and
therefore Birman--Kuroda theorem applies \cite[Sec.XI.3]{RS}. What is
more interesting, however, is how the scattering matrix looks like.
To find it we employ the relation (\ref{decomposition}) in
combination with (\ref{lambda}) changing slightly the notation: to
any $\,\psi\in D(H(\alpha,\vec a))\,$ and a nonreal $\,z\,$ there is
a unique decomposition
$$
\psi(\vec x)\,=\, \psi_z(\vec x)\,+\, {1 \over \alpha-\xi(\vec
a,z)}\, G_0(\vec x,\vec a;z) \psi_z(\vec a)
$$
with $\,\psi_z\in D(H_0)\,$ and $\,(H(\alpha,\vec a)-z)\psi=
(H_0-z)\psi_z\,$. If we choose
   \begin{equation} \label{scattering Ansatz}
\psi_z^{\eps}\,:=\, e^{ik_n(z)x-\eps x^2} \chi_n(y)
   \end{equation}
for $\,\psi_z\,$, the corresponding $\,\psi=: \psi^\eps\,$ belongs to
$\,D(H(\alpha,\vec a))\,$ for all $\,\eps>0\,$ and
$$
((H(\alpha,\vec a)-z)\psi^{\eps})(\vec x)\,=\, 2\eps(2\eps x^2-
1-2ik_n(z)x) \psi_z^{\eps}(\vec x)\,.
$$
Since the \rhs converges in $\,L^2\,$ sense as $\,z\,$ approaches the
real line, $\,\psi^{\eps}\in D(H(\alpha,\vec a))\,$ for $\,z\in
[1,\infty)\,$ and the last relation holds again. Of course,
$\,\psi^\eps\,$ ceases to be square integrable as $\,\eps\to 0+\,$,
but the pointwise limit exists and
   \begin{equation} \label{generalized eigenfunction}
\psi(\vec x)\,=\, e^{ik_n(z)x} \chi_n(y)\,+\, {e^{ik_n(z)a} \over
\alpha-\xi(\vec a,z)}\, G_0(\vec x,\vec a;z) \chi_n(b)\,.
   \end{equation}
This function is locally square integrable, satisfies the appropriate
boundary conditions and solves $\,(H(\alpha,\vec a)-z)\psi=0\,$ as a
differential equation. Substituting for $\,G_0\,$ we obtain easily
the reflection and transmission amplitudes, $\,r_{nm}(z)\,$ and
$\,t_{nm}(z)\,$, respectively, for the case when the incident wave
corresponds to the $\,n$--th transverse mode and the scattered one
appears in the $\,m$--th mode; they are given by the relation
   \begin{equation} \label{rt}
(t_{nm}(z)- \delta_{nm})\, e^{i(k_m-k_n)a}\,=\, r_{nm}(z)\, e^{-i(k_n
+k_m)a}\,=\, {i\over\pi}\, {\sin(nb) \sin(mb)\over k_m(z) (\alpha-
\xi(\vec a,z))}\,.
   \end{equation}
It is necessary to assume, of course, that both the involved channels
are open, \ie, $\,z> \max(n^2,m^2)\,$.

Since the point interaction gives rise to a rank--one perturbation in
the resolvent, the Kato--Birman theory yields also the completeness
of the wave operators, \ie, unitarity of the S--matrix. In
particular, the ``diagonal'' part of the unitarity condition can be
expressed in terms of the transmission and reflection coefficients as
   \begin{equation} \label{unitarity}
\sum_{m=1}^{[\sqrt z]} k_m (|t_{nm}|^2+|r_{nm}|^2) \,=\, k_n\,,
   \end{equation}
and has an obvious meaning of conservation of probability flow
(taking into account different particle velocities in different
channels). It is not difficult in the present case to check the
identity (\ref{unitarity}) directly. Since $\,t_{nn}= 1+r_{nn}
e^{-2ik_n a}\,$, it reduces to
$$
k_n\, \re\left( r_{nn} e^{-2ik_n a}\right) \,+\, 2\,
\sum_{m=1}^{[\sqrt z]} k_m|r_{nm}|^2\,=\,0 \;;
$$
substituting from (\ref{rt}) and multiplying both sides by $\,|\alpha
-\xi(\vec a,z)|^2\,$, we arrive at the relation
$$
\im \xi(\vec a,z)\,=\, {1\over\pi}\, \sum_{m=1}^{[\sqrt{z}]}\,
{\sin^2(mb)\over k_m(z)}\,,
$$
which is nothing else than (\ref{im part}); recall that the closed
channels do not contribute to the imaginary part of $\,\xi(\vec
a,z)\,$. In the same way one can check the remaining parts of the
unitarity condition which read
$$
\sum_{m=1}^{[\sqrt z]} k_m (t_{nm}\overline t_{sm}+r_{nm}\overline
r_{sm}) \,=\,\delta_{ns} k_n
$$
for $\,n\ne s\,$ and
$$
\sum_{m=1}^{[\sqrt z]} k_m \left(t_{nm}\overline r_{sm} e^{2ik_ma}
+r_{nm}\overline t_{sm}e^{-2ik_ma}\right) \,=\,0 \,,
$$
since the ``elementary block'' of the S--matrix describing the
transitions between the $\,n$--th and $\,m$--th channel is of the
form
   \begin{equation} \label{S block}
S_{nm}\,=\, \sqrt{k_m\over k_n}\, \left( \matrix{ t_{nm} & r_{nm} \cr
\tilde r_{nm} & \tilde t_{nm} } \right)\,,
   \end{equation}
where the tilded quantities in the second row are obtained by
changing the sign of $\,a\,$, \ie, $\,\tilde r_{nm}:= r_{nm}
e^{-2i(k_n+k_m)a}\,$ \etc

\subsection{Resonances}

Every square root in the definition of the channel momentum
$\,k_n(z):= \sqrt{z-n^2}\,$ gives rise to a cut; hence the free
resolvent kernel (\ref{free kernel}) as well as other quantities
derived from it are in general multivalued with infinitely sheeted
Riemann surfaces. In particular, the pole condition (\ref{pole
condition}) may have solutions on the other (nonphysical) sheets
which produce resonances.

For simplicity we introduce $\,q_n(z):= \sqrt{z-n^2}\,$ which takes
values always in the upper complex halfplane or at the positive real
halfline. On the $\,(N+1)$--th sheet we have then
$$
k_n(z)\,:=\, \left\lbrace\: \begin{array}{ccc} -q_n(z) \qquad & \dots
\qquad & n\le N \\ q_n(z) \qquad & \dots \qquad & n>N \end{array}
\right.
$$
Denoting by $\,\xi_N(\vec a,\cdot)\,$ the corresponding branch of the
function $\,\xi(\vec a,\cdot)\,$ we can write the pole condition
(\ref{pole condition}) explicitly as
   \begin{eqnarray} \label{resonance condition}
\alpha- \re\xi_N(\vec a,z) &\!=\!& \alpha +\,{1\over\pi}\,
\sum_{n=1}^N \left( -\im\left( 1\over q_n(z)\right) \sin^2(nb)+
\,{1\over 2n} \right) \nonumber \\ &&
+\,{1\over\pi}\, \sum_{n=N+1}^{\infty} \left( \im\left( 1\over
q_n(z)\right) \sin^2(nb)+ \,{1\over 2n} \right) \,=\,0\,, \nonumber
\\ \\ -\,\im\xi_N(\vec a,z) &\!=\!& {1\over\pi}\,\sum_{n=1}^N
\re\left( 1\over q_n(z)\right) \sin^2(nb) \nonumber \\
&& -\,{1\over\pi}\,\sum_{n=N+1}^{\infty} \re\left( 1\over
q_n(z)\right) \sin^2(nb) \,=\,0 \,.\nonumber
   \end{eqnarray}
For a weak coupling, there is generically one resonance pole close to
each threshold, with exception of the lowest one; the resonance is
absent if $\,(N\!+\!1)b/\pi\,$ is integer so that the incident wave
has a node at the impurity and does not feel the perturbation. To
show this more explicitly, let us rewrite $\,\xi_N(\vec a,z)=
\xi^N(\vec a,z)+ \tilde\xi^N(\vec a,z)\,$, where the two terms sum
contributions from the evanescent and propagating modes,
$\,n\ge N\!+\!1\,$ and $\,1\le n\le N\,$, respectively. Suppose we
switch the latter on with the help of an additional parameter, \ie,
we look for solutions of the condition
   \begin{equation} \label{auxiliary resonance}
F(z,\eta)\,:=\, \alpha-\xi^N(\vec a,z)-\eta\tilde\xi^N(\vec
a,z)\,=\,0\,.
   \end{equation}
If $\,\eta=0\,$, we can repeat the argument of Sec.~2.5: we find that
the function $\,\xi^N(\vec a,\cdot)\,$ is monotonously growing
between $\,-\infty\,$ and $\,\infty\,$ when $\,z\,$ runs over the
interval $\,(-\infty, (N\!+\!1)^2)\;$; hence to a given $\,\alpha\,$
there is just one $\,z_N^0(\alpha)\,$ such that
$\,F(z_N^0(\alpha),0)=0\,$. Moreover, the leading behavior of
$\,\xi^N(\vec a,\cdot)\,$ as $\,z\to (N\!+\!1)^2\!-\,$ is again given
by a single term, so the ``eigenvalues'' behave as
   \begin{equation} \label{auxiliary pole}
z_N^0(\alpha)\,=\, (N\!+\!1)^2- \left( \sin^2((N\!+\!1)b)\over \pi\alpha
\right)^2 +\OO(\alpha^{-3})
   \end{equation}
in the weak--coupling case, $\,\alpha\to\infty\,$.

For a nonzero $\,\eta\,$ the condition (\ref{auxiliary resonance})
can be solved perturbatively by means of the implicit--function
theorem. We have
   \begin{eqnarray*}
\left.{\partial F\over\partial\eta}\right|_{(z_N^0(\alpha),0)}
\!&=&\! -\tilde\xi^N(z_N^0(\alpha))\,=\, {i\over\pi}\, \sum_{n=1}^N\,
\left( {\sin^2(nb)\over \sqrt{(N\!+\!1)^2\!-n^2}}\,+\,{1\over 2in}
\right)\,+\, \OO(\alpha^{-1})\,, \\ \\
\left.{\partial F\over\partial z}\right|_{(z_N^0(\alpha),0)}
\!&=&\! {i\over2\pi}\, \sum_{n=N+1}^{\infty}\,
{\sin^2(nb)\over q_N(z_N^0(\alpha))^3} \,=\, {\pi^3\alpha^3 \over
|\sin((N\!+\!1)b))|}\,+\, \OO(\alpha^0)\;;
   \end{eqnarray*}
dividing these quantities we obtain $\,-(\partial z_N^{\eta}/
\partial\eta)_{\eta=0}\,$. Moreover, the remainder term coming from
$\,\partial^2 z_N^{\eta}/\partial\eta^2\,$ is $\,\OO(\alpha^{-4})\,$,
so for sufficiently large positive $\,\alpha\,$ we may use the
expansion up to $\,\eta=1\,$, obtaining an asymptotic formula for the
resonance--pole position,
   \begin{equation} \label{weak pole position}
z_N(\alpha)\,=\, (N\!+\!1)^2-\, 2i\, {|\sin((N\!+\!1)b)|\over
\pi^3\alpha^3} \sum_{n=1}^N\,\left( {\sin^2(nb)\over
\sqrt{(N\!+\!1)^2\!-n^2}}\,+\,{1\over 2in} \right)\,+\,
\OO(\alpha^{-4})\;;
   \end{equation}
the real shift appearing in (\ref{auxiliary pole}) has been absorbed
here into the error term. In particular, the resonance width behaves
in the weak--coupling case as
   \begin{equation} \label{weak width}
\Gamma_N(\alpha)\,:=\, -2\,\im z_N(\alpha)\,=\, 4\,
{|\sin((N\!+\!1)b)|\over \pi^3\alpha^3} \sum_{n=1}^N\,
{\sin^2(nb)\over \sqrt{(N\!+\!1)^2\!-n^2}}\,+\,
\OO(\alpha^{-4})\,.
   \end{equation}
If the coupling is not weak, resonance poles can be found from a
numerical solution of the conditions (\ref{resonance condition}); an
example is shown on Figure~4. The pole trajectories can be followed
for all values of $\,\alpha\,$, however, only weak perturbations
produce a substantial resonance scattering effect because the pole
residues decrease rapidly with the coupling strength; this also is 
shown on Figure~4.
   \begin{figure} \label{1st resonance}
   \caption{Pole trajectory in the $\,z$--plane.}
   \end{figure}

\section{A finite number of point interactions} \label{finite number}

\setcounter{equation}{0}

Our next aim is to extend the above results to the situation when
there is a finite number $\,N\,$ of point interactions in the strip
$\,\Omega\,$. We suppose that their positions are $\,\vec a_j:=
(a_j,b_j)\,$ and the coupling constants $\,\alpha_j\;$; for the sake
of brevity we denote $\,\vec a:=\{\vec a_1,\dots,\vec a_N\}\,$ and
$\,\alpha:= \{\alpha_1,\cdots,\alpha_N\}\,$. The corresponding
Hamiltonian $\,H(\alpha,\vec a)\,$ is again given by (\ref{one-center
perturbation}) with the boundary conditions (\ref{bc}) replaced by
   \begin{equation} \label{bc_N}
L_1(\psi,\vec a_j)\,+\, 2\pi\alpha_j L_0(\psi,\vec a_j)\,=\,0\,,
\qquad j\,=\,1,\dots,N\,.
   \end{equation}
As in the one--center case, any of these point interactions may be
switched off by means of the limit $\,\alpha_j\to\infty\,$.

\subsection{The resolvent}

We shall again start from derivation of the resolvent. Since the
deficiency indices of the initial symmetric operator are now
$\,(N,N)\,$, a natural Ansatz is
   \begin{equation} \label{Krein_N}
(H(\alpha,\vec a)\!-\!z)^{-1}(\vec x_1,\vec x_2)\,=\, G_0(\vec
x_1,\vec x_2;z) \,+\,\sum_{j,k=1}^N \lambda_{jk} G_0(\vec x_1,\vec
a_j;z) G_0(\vec a_k,\vec x_2;z)\,,
   \end{equation}
where $\,G_0(\cdot,\cdot;z)\,$ is the free--resolvent kernel
(\ref{free kernel}). Functions $\,\psi\in D(H(\alpha,\vec a))\,$ and
$\,\psi_0\in D(H_0)\,$ introduced as above are then related by
   \begin{equation} \label{decomposition_N}
\psi(\vec x)\,=\, \psi_0(\vec x)\,+\, \sum_{j,k=1}^N \lambda_{jk}
G_0(\vec x,\vec a_j;z)\, \psi_0(\vec a_k)
   \end{equation}
and one can compute easily the boundary values at the interaction
support,
   \begin{eqnarray*}
L_0(\psi,\vec a_m) &\!=\!& -\, \sum_{j,k=1}^N {\lambda_{jk} \over
2\pi}\, \delta_{jm}\, \psi_0(\vec a_k)\,, \\ \\
L_1(\psi,\vec a_m) &\!=\!& -\,\psi_0(\vec a_m)\,+\,  \sum_{j,k=1}^N
\lambda_{jk}\, \Biggl( {i\over\pi}\, \delta_{jm}\,
\sum_{n=1}^{\infty}\, \left(\, {\sin^2(nb_j)\over k_n(z)}\,-\,
{1\over 2in} \,\right)\, \\ && +\, (1- \delta_{jm}) G_0(\vec a_j,\vec
a_m;z)\, \Biggr) \psi_0(\vec a_k)\,.
   \end{eqnarray*}
Substituting this into (\ref{bc_N}) and using the fact that
$\,\psi_0(\vec a_k)\,$ are independent quantities, we find
   \begin{equation} \label{lambda_N}
\lambda(\alpha,\vec a;z)\,=\, \Lambda(\alpha,\vec a;z)^{-1}\,,
   \end{equation}
where $\,\Lambda(\alpha,\vec a;z)\,$ is now an $\,N\times N\,$ matrix
given by
   \begin{eqnarray} \label{Lambda_N}
\Lambda_{jj} &\!=\!& \alpha_j\,-\,{i\over\pi}\, \sum_{n=1}^{\infty}\,
\left(\, {\sin^2(nb_j)\over k_n(z)}\,-\, {1\over 2in} \,\right)\,,
\nonumber \\ \\
\Lambda_{jm} &\!=\!& -G_0(\vec a_j,\vec a_m;z) \,=\, -\,{i\over\pi}\,
\sum_{n=1}^{\infty} {e^{ik_n(z)|a_j-a_m|}\over k_n(z)}\,
\sin(nb_j) \sin(nb_m)\,, \quad j\,\ne m\;; \nonumber
   \end{eqnarray}
we shall again drop the arguments occassionally if they are clear
from the context.  Notice that the scaling argument of the
Section~2.3 can be easily adapted to the present situation; it shows
that changing $\,d\,$ is equivalent to the {\em simultaneous} shift
of all the coupling constants $\,\alpha_j\,$ on $\,-\,{1\over 2\pi}\,
\ln\left(\pi\over d\right)\,$ in the rescaled strip.

\subsection{The discrete spectrum}

A finite number of point interactions represents a finite--rank
perturbation in the resolvent, and therefore the conclusion of
Sec.2.5 about the essential spectrum remains valid, $\,\sigma_{ess}
(H(\alpha,\vec a))= [1,\infty)\,$. The discrete spectrum is again
determined by poles of the resolvent coming from the coefficients
$\,\lambda_{jk}\,$ in (\ref{Krein_N}); they are given by the
condition
   \begin{equation} \label{pole_N}
\det \Lambda(\alpha,\vec a,z)\,=\,0\,.
   \end{equation}
Comparing with the case $\,N=1\,$, it is now slightly more complicated
to determine the eigenfunctions. One can adapt the argument from
\cite[Sec.II.1]{AGHH}. Suppose that $\,H:=H(\alpha,\vec a)\,$
satisfies $\,H\varphi= z\varphi\,$ for some $\,z\in\R\,$. We  pick an
arbitrary $\,z'\in\rho(H)\;$; then in analogy with
(\ref{decomposition_N}) there is a vector $\,\psi_0\in D(H_0)\,$
which allows us to express the eigenvector $\,\varphi\,$ as
   \begin{equation} \label{eigenvector decomposition}
\varphi\,=\, \psi_0+\, \sum_{j=1}^N d_j G_0(\cdot,\vec a_j;z')\,,
   \end{equation}
where the coefficients are given by $\,d_j:= \sum_{k=1}^N
(\Lambda(z')^{-1})_{jk} \psi_0(\vec a_k)\,$ and the relations
$$
(H_0-z')\psi_0\,=\, (H-z')\varphi\,=\, (z-z')\varphi
$$
are valid. Applying $\,(H_0-z')^{-1}\,$ to the last identity we
obtain
$$
\psi_0\,=\, (z-z') \left\lbrack (H_0-z')^{-1}\psi_0+\, \sum_{j=1}^N
d_j (H_0-z')^{-1} G_0(\cdot,\vec a_j;z') \right\rbrack\,,
$$
and this in turn yields
   \begin{equation} \label{identity 1}
(H_0-z)\psi_0\,=\, (z-z')\, \sum_{j=1}^N d_j G_0(\cdot,\vec a_j;z')
\,.
   \end{equation}
If $\,z<1\,$, the resolvent $\,(H_0-z)^{-1}\,$ exists and may be
applied to both sides of the last relation giving
   \begin{equation} \label{identity 2}
\psi_0\,=\, \sum_{j=1}^N d_j \left( G_0(\cdot,\vec a_j;z)-
G_0(\cdot,\vec a_j;z') \right) \;;
   \end{equation}
we have employed here the first resolvent identity. Substituting into
(\ref{eigenvector decomposition}) we get an expression for the
(non--normalized) eigenfunction
   \begin{equation} \label{eigenfunction_N}
\varphi(\vec x)\,=\, \sum_{j=1}^N d_j G_0(\vec x,\vec a_j;z)\,.
   \end{equation}
To determine the coefficients, we notice that the relation
(\ref{identity 2}) together with (\ref{Lambda_N}) gives
$$
\psi_0(\vec a_j)\,=\, \sum_{m=1}^N \left( \Lambda(z')_{jm}-
\Lambda(z)_{jm} \right) d_m \;;
$$
on the other hand, the above mentioned expression of $\,d_j\,$ is
equivalent to $\,\psi_0(\vec a_j)\,=\, \sum_{m=1}^N \Lambda(z')_{jm}
d_m\,$. Hence
   \begin{equation}\label{eigenfunction condition}
\sum_{m=1}^N \Lambda(z)_{jm} d_m\,=\,0\,,
   \end{equation}
\ie, $\,d:= (d_1,\dots,d_N)\,$ is an eigenvector of
$\,\Lambda(\alpha,\vec a,z)\,$ corresponding to zero eigenvalue.
Inverting the argument as in \cite{AGHH} one can check that any
solution to (\ref{eigenfunction condition}) determines an eigenvector
of $\,H(\alpha,\vec a)\,$.

The next question concerns the existence of solutions to
Eqs(\ref{pole_N}) and (\ref{eigenfunction condition}). For
convenience we introduce again $\,\kappa_n= -ik_n(z)=
\sqrt{n^2-z}\,$ which is positive for $\,z<1\,$. We have then
$$
\Lambda_{jj} \,=\, \alpha_j\,-\,{1\over\pi}\, \sum_{n=1}^{\infty}\,
\left(\, {\sin^2(nb_j)\over \kappa_n(z)}\,-\, {1\over 2n}
\,\right)\,,
$$
and an analogous expression for the nondiagonal elements. If $\,z\to
-\infty\,$ the matrix behaves as
$$
\Lambda(\alpha,\vec a,z)\,=\, {1\over 4\pi}\, \ln\left( -\,{z\over
4}\right)\,I \,+\,\OO(z^0)\,,
$$
so all its eigenvalues tend to $\,+\infty\,$ as $\,z\to-\infty\,$. On
the other hand, for $\,z\to 1-\,$ we have
$$
\Lambda(\alpha,\vec a,z)\,=\, -\,{1\over \pi\sqrt{1-z}}\,
M_1\,+\,\OO(1)\,,
$$
where $\,M_1:= (\sin b_j\, \sin b_m)_{j,m=1}^N\,$. This matrix has,
in particular, an eigenvector $\,(\sin b_1,\dots, \sin b_N)\,$
corresponding to the {\em positive} eigenvalue $\, \sum_{j=1}^N
\sin^2 b_j\,$, and therefore one of the eigenvalues of
$\,\Lambda(\alpha,\vec a,z)\,$ tends to $\,-\infty\,$ as $\,z\to
1-\,$. The elements of $\,\Lambda(\alpha,\vec a,z)\,$ are continuous
functions of $\,z\,$, hence the same is true for its eigenvalues. It
follows that there is an eigenvalue which crosses zero, \ie,
$\,H(\alpha,\vec a)\,$ has {\em at least one eigenvalue.}

In fact a stronger claim can be made concerning the eigenvalues of
$\,\Lambda(\alpha,\vec a,z)\,$. By a straightforward differentiation
we find
$$
{d\over dz}\, \Lambda(z)_{jm}\,=\, -\,{1\over 2\pi}\,
\sum_{n=1}^{\infty} {e^{-|a_j-a_m|\sqrt{n^2-z}} \over
(n^2-z)^{3/2}}\, \left( 1+|a_j-a_m|\sqrt{n^2-z}\right)\,
\sin(nb_j)\sin(nb_m)\,. 
$$
The matrix function $\,\Lambda(\cdot)\,$ is monotonous if for any
$\,c\in\C^N\,$ the quantity
   \begin{eqnarray*}
{d\over dz}\, (c,\Lambda(z)c) &\!=\!&  -\,{1\over 2\pi}\,
\sum_{n=1}^{\infty} (n^2-z)^{-3/2}\, \sum_{j,m=1}^N
\overline{c_j\sin(nb_j)}\, c_m\sin(nb_m) \\ && \times\,
e^{-|a_j-a_m|\sqrt{n^2-z}}  \left( 1+|a_j-a_m|\sqrt{n^2-z}\right)
   \end{eqnarray*}
has a definite sign (is nonpositive in our case). The expression on
the right hand side tells us that this is true provided the function
$\,f:\: f(x)= e^{-\kappa|x|} (1+\kappa|x|)\,$ is of positive type
for any $\,\kappa>0\,$. The last named property follows from the
identity
$$
(1+\kappa|x|)\, e^{-\kappa|x|} \,=\, {2\kappa^3\over \pi}\,
\int_{\R} {e^{ipx}\over (p^2+\kappa^2)^2}\, dp\,,
$$
because by Bochner's theorem \cite[Sec.IX.2]{RS} a function is of
positive type \Iff its Fourier image is positive. In fact, since
the measure in the last integral is pointwise positive, $\,{d\over
dz}\, \Lambda(z)\,$ is even strictly positive; it means that all the
eigenvalues of $\,\Lambda(\alpha,\vec a;z)\,$ are decreasing
functions of $\,z\,$.

The most important consequence of this fact is the {\em nondegeneracy
of the ground state.} To prove it one has to check that the lowest
eigenvalue of $\,\Lambda(z)\,$ is simple for any $\,z\in
(-\infty,1)\,$, which is equivalent to the claim that the matrix
semigroup $\, \lbrace\, e^{-t\Lambda(z)} :\, t\ge 0\,\rbrace\,$ is
positivity preserving \cite[Sec.XIII.12]{RS}. The last property is
ensured if all the nondiagonal elements of $\,\Lambda(z)\,$ are
negative; we have $\,\Lambda(z)_{jm}= -G_0(\vec a_j,\vec a_m;z)\,$ by
(\ref{Lambda_N}) so the desired result follows from the positivity of
the free--resolvent kernel. The coefficients may be therefore chosen
of the same sign for the ground state; in fact, as positive because
$\,d_{j_0}=0\,$ would mean that the eigenfunction
(\ref{eigenfunction_N}) is smooth at $\,\vec x=\vec a_{j_0}\,$ so the
corresponding interaction is absent, $\,\alpha_{j_0}= \infty\,$.
   \begin{figure} \label{multiple point eigenfunctions}
   \caption{Eigenfunctions for three point perturbations. (a) The
ground state. (b) The second excited state.}
   \end{figure}

\subsection{Embedded eigenvalues}

Let us now ask whether the continuous spectrum of $\,H:=H(\alpha,
\vec a)\,$ may contain eigenvalues if $\,N>1\,$. Suppose that
$\,H\varphi= z\varphi\,$ for some $\,z>1\,$. We employ again the
expression (\ref{eigenvector decomposition}) for the eigenvector and
write $\,\psi_0\,$ as a series, $\,\psi_0(\vec x)=\,
\sum_{n=1}^{\infty} g_n(x) \chi_n(y)\,$, with the coefficient
functions $\,g_n\in L^2(\R)\,$. Substituting this into (\ref{identity
1}) and using the fact that $\,\{ \chi_n\}\,$ is an orthonormal basis
in $\,L^2(0,d)\,$, we obtain the following system of equations,
$$
-g''_n(x) -k_n(z)^2 g_n(x)\,=\, {i\over 2}\, (z-z')\, \sum_{j=1}^N
d_j \chi_n(b_j)\, {e^{ik_n(z')|x-a_j|}\over k_n(z')}
$$
for $\,n=1,2,\dots\;$; the Fourier--Plancherel operator transforms it
to
   \begin{equation} \label{g hat}
(p^2-z+n^2)\hat g_n(p)\,=\, {z-z'\over 2\pi}\, \sum_{j=1}^N
d_j \chi_n(b_j)\, {e^{-ipa_j}\over p^2-z'+n^2} \,.
   \end{equation}
If $\,g_n\in L^2\,$ the same is true for $\,\hat g_n\;$; this is
impossible if
   \begin{equation} \label{embedded ev restriction}
z\,>\,n^2
   \end{equation}
and the \rhs of (\ref{g hat}) is nonzero at $\,\pm p_n\,$, where
$\,p_n:= \sqrt{n^2\!-z}\,$, since $\,\hat g_n^2\,$ would have then a
nonintegrable singularity. It is clear that it is the factor
$\,p^2\!-z+n^2\,$ which matters, because $\,z'\,$ belongs to
$\,\rho(H)\,$ by assumption.

Hence the \rhs of (\ref{g hat}) has to be zero; we want to conclude
that $\,\hat g_n=0\,$. If $\,N>1\,$ and $\,a_j\,$ are not the same,
it might happen that the \rhs is not zero identically. However, we
can choose a common phase factor to be put in front of the sum; then
the square integrability requires
$$
\sum_{j=1}^N d_j\chi_n(b_j)\, e^{\mp ip_n(a_j-a)}\,=\,0
$$
for an arbitrary $\,a\,$. If all the $\,a_j\,$ are mutually different
(mod $\,2\pi p_n^{-1}$) it follows that $\,d_j\chi_n(b_j)=0\,$ for
each $\,j\,$. On the other hand, if some of them coincide we find
$\,\sum_j d_j\chi_n(b_j)=0\,$ where the index runs through the values
with the same longitudinal coordinate $\,a_j\,$, and therefore
$\,\hat g_n=0\,$ again, \ie, $\,\hat g_n\,$ may be nonzero at most if
some $\,a_j\,$ differ by multiples of $\,2\pi p_n^{-1}$.

Consider now an arbitrary $\,g\in L^2(\R)\,$ and $\,n\,$ satisfying
the condition (\ref{embedded ev restriction}). Using
(\ref{eigenvector decomposition}) and (\ref{free kernel}) we find
$$
(g\chi_n,\varphi) \,=\, (\hat g,\hat g_n)\,+\, {i\over 2k_n(z')}\,
\sum_{j=1}^N d_j \chi_n(b_j) (g, e^{ik_n(z')|\cdot-a_j|}) \,,
$$
where the inner product on the \rhs refers to $\,L^2(\R)\,$ and
$\,\hat g_n\,$ in the first term can be expressed by (\ref{g hat}).
If $\,d_j\chi_n(b_j)=0\,$ for each $\,j\,$, the right hand side is
zero. In the exceptional case mentioned above we use the fact that
the \lhs is independent of $\,z'\,$. The explicit expression for
$\,d_j\,$ together with the asymptotic behavior of $\,\Lambda\,$ show
that $\,d_j\to 0\,$ as $\,z'\to -\infty\,$; the same is true for the
inner product in the second term as well as for $\,(\hat g,\hat
g_n)\,$; together we find $\,(g\chi_n,\varphi)=0\,$ again. We
conclude that {\em $\,z>1\,$ cannot be an eigenvalue corresponding to
an eigenvector from the subspace $\,\bigoplus_{n=1}^{[\sqrt z]}
L^2(\R)\otimes \{\chi_n\}\,$.}

On the other hand, the condition (\ref{embedded ev restriction}) in
the above argument is crucial; in the case $\,N>1\,$ the operator
$\,H(\alpha,\vec a)\,$ {\em can have embedded eigenvalues} with
eigenfunctions in the orthogonal complement of the mentioned
subspace. As the simplest example, consider a pair of point
perturbations with the {\em same} coupling constant $\,\alpha\,$
placed at $\,\vec a_1:= (0,b)\,$ and $\,\vec a_1:= (0,\pi\!-\!b)\,$.
The eigenvalue problem can be divided into the part symmetric and
antisymmetric with respect to the strip axis, the antisymmetric part
being obtained by scaling of the single--center problem with $\,\vec
a:= (0,2b)\,$ and coupling constant $\,\alpha-\,{1\over 2\pi}\ln
2\,$. The scaled eigenvalue tends to $\,4\,$ as $\,\alpha\to
\infty\,$, hence it is embedded in the continuum for all $\,\alpha\,$
large enough. This is illustrated on Figure~6.
   \begin{figure} \label{embedded ev}
   \caption{An embedded eigenvalue due to symmetry.}
   \end{figure}
In the same way, one can construct other examples of embedded
eigenvalues. Their common feature is the existence of a symmetry
which prevents the (energetically allowed) decay of the eigenstate; a
violation of the symmetry turns these eigenvalues into resonances.
Recall also that embedded eigenvalues due to symmetric obstacles in
{\em Neumann} waveguides have been treated recently in \cite{ELV}.

\subsection{The limits of strong and weak coupling}

If all the point interactions under consideration are strong, \ie,
having the $\,\alpha_j$'s large negative, one naturally expects the
corresponding bound states to be strongly localized and weakly
influenced both by the other perturbations and by the boundary. To
show that this is indeed the case, let us write the matrix
$\,\Lambda(\alpha, \vec a;z)\,$ in the form
$$
\Lambda(z)\,=\, \left(\left( \alpha_j+\,{1\over 4\pi}\, \ln\left(
-\,{z\over 4}\right)\right) \delta_{jk}\right) \left\lbrack I\,+\,
\left(\left( \alpha_j+\,{1\over 4\pi}\, \ln\left( -\,{z\over
4}\right)\right) \delta_{jk}\right)^{-1} \tilde\Lambda(z)
\right\rbrack\,,
$$
where $\,\tilde\Lambda(z)\,$ is the remainder matrix, which is
independent of $\,\alpha\,$ and has a bounded norm as $\,z\to
-\infty\,$. Given a finite energy interval $\,I\subset (-\infty,1)\,$
one can always choose the$\,\alpha_j$'s large enough negative so that
the two matrices in the above product are regular, and no eigenvalues
are contained in $\,I\,$. Consequently, the roots of Eq.(\ref{pole
condition}) in the strong--coupling limit are situated in the region
where $\,\Lambda(z)\,$ is dominated by the diagonal part; then there
are exactly $\,N\,$ eigenvalues, including a possible degeneracy, and
   \begin{equation} \label{strong asymptotics}
\epsilon_j(\alpha,\vec a) \,\approx\, 4\,e^{-4\pi\alpha_j} \qquad
{\rm as} \qquad \max_{1\le j\le N}\alpha_j \to-\infty\,.
   \end{equation}
One can conjecture that the asymptotics is in fact the same as in
(\ref{energy asymptotics}). The corresponding eigenfunctions coincide
in the leading order with the strongly coupled one--center
eigenfunctions of Section~2.5.

For a weak coupling the situation is different. We again restrict our
attention to the case when {\em all} the interactions are weak, \ie,
the corresponding $\,\alpha_j$'s are large positive. We denote $\,A:=
{\rm diag}(\alpha_1,\dots,\alpha_N)\,$ and use the decomposition
$$
\Lambda(z)\,=\, \left( A\,-\, {1\over \pi\sqrt{1-z}}\,M_1 \right)
\left\lbrack I\,+\, \left( A\,-\, {1\over \pi\sqrt{1-z}}\,M_1
\right)^{-1} \tilde\Lambda(z)\right\rbrack\,,
$$
where $\,\tilde\Lambda(z)\,$ is a remainder independent of
$\,\alpha\,$, whose norm is bounded as $\,z\to 1-\,$. In the same way
as above, to a given $\,z_0<1\,$ one can always choose the
$\,\alpha_j$'s sufficiently large so that the two matrices are
regular and no eigenvalues are contained in $\,(-\infty,z_0]\,$. In
distinction to the previous case, however, $\,\Lambda(z)\,$ is
asymptotically not a multiple of the unit matrix but rather a
rank--one operator on $\,\C^N\,$ as we have seen in the previous
section. Hence only one eigenvalue of $\,H(\alpha,\vec a)\,$ can
approach the threshold of the continuous spectrum; in combination
with the above argument this means that if all the point interactions
are weak enough, $\,H(\alpha,\vec a)\,$ has {\em a single bound
state.}

This is analogous to the behavior of interactions whose support has a
nonzero measure. Recall that a two--dimensional Schr\"odinger
operator with a weak potential $\,V\,$ has always one bound state
provided $\,V\,$ is attractive in the mean and satisfies some decay
restrictions \cite{Si}, and the same is true in a tube, whatever
dimension its cross section is, provided the potential is replaced by
its projection on the lowest transverse mode \cite{Ex}.

To find the corresponding bound state in our case, we have to solve
the spectral problem for $\,M_1+\eta A\,$, where $\,\eta:= -\pi
\sqrt{1-z}\;$; we are interested in the eigenvalue that approaches
$\,\sum_{j=1}^N \sin^2 b_j\,$ as $\,z\to 1-\,$. An elementary
perturbative argument yields the expression
$$
\sum_{j=1}^N \sin^2 b_j\,-\,\pi\sqrt{1-z}\, {\sum_{j=1}^N
\alpha_j\sin^2 b_j \over \sum_{j=1}^N \sin^2 b_j}\,+\, \OO(\eta^2)\;;
$$
putting it equal to zero, we obtain the energy of the weakly bound
state,
   \begin{equation} \label{weak asymptotics}
\epsilon(\alpha,\vec a)\,\approx\, 1\,-\, \left( {\left( \sum_{j=1}^N
\sin^2 b_j \right)^2 \over \pi\, \sum_{j=1}^N \alpha_j\sin^2 b_j}
\right)^2
   \end{equation}
as $\,\min_{1\le j\le N} \alpha_j\to\infty\,$. Since the range of
$\,M_1\,$ is spanned by $\,(\sin b_1,\dots,\sin b_N)\,$, the
corresponding asymptotic expression for the eigenfunction is
   \begin{eqnarray*}
\psi(x;\alpha,\vec a) &\!\approx\!& \sin y \: {\sum_{k=1}^N
\alpha_k\sin^2 b_k \over \left( \sum_{k=1}^N \sin^2 b_k \right)^2}\:
\sum_{j=1}^N e^{-\sqrt{1-\epsilon}|x-a_j|} \sin^2 b_j \\ \\ &&
+\, \sum_{n=2}^{\infty}\, \sin(ny)\: \sum_{j=1}^N
{e^{-\sqrt{n^2-1}|x-a_j|} \over \sqrt{n^2-1}}\, \sin(nb_j) \sin
b_j\,.
   \end{eqnarray*}
The leading term is again a product of $\,\chi_1(y)\,$ with a linear
combination of the eigenfunctions of one--dimensional point
interactions placed at $\,a_j,\; j=1,\dots,N\,$.

\subsection{More about bound states}

The spectral condition (\ref{pole_N}) makes it possible to derive
also other properties of the discrete spectrum for the $\,N$--center
Hamiltonian. We have seen that it has $\,N\,$ bound states and just
one bound state in the strong-- and weak--coupling limits,
respectively. One may ask, more generally, what are the regions in
the space $\,\R^N\,$ of coupling constants, where the dimension of
the discrete spectrum is $\,n=1,2,\dots,N\,$.

First consider the two--center case. As we already mentioned $\,\det
\Lambda (\alpha, \vec a;z)\,$ tends to $\,+\infty\,$ as $\,z \to
-\infty\,$. On the other hand, the behavior of this quantity as $\,z
\to 1-\,$ depends on the parameters $\,\alpha_j,\,\vec a_j,\;
j=1,2\;$; the limit may be either $\,+\infty\,$ or $\,-\infty\,$,
\ie, the equation $\,\det\,\Lambda (\alpha,\vec a;z)=0\,$ can have
one or two roots. The corresponding areas in the $\,(\alpha_1,
\alpha_2)$--plane  are divided by the line
$\,P_2(\alpha_1,\alpha_2)=0\,$, where
$$
P_2(\alpha_1,\alpha_2)\,:=\,\alpha_1\,\sin^{2}b_2\,+\,\alpha_2\,
\sin^{2}b_1\,+\,{\sin^{2}b_1\!+\sin^{2}b_2\over2\pi}\,-\,{2\over\pi}\,
|a_1\!-a_2|\, \sin^{2}b_1\,\sin^{2}b_2\;;
$$
there is only one bound state if $\,P_2(\alpha_2,\alpha_2)\,\ge\,0\,$
and two in the opposite case. Inspecting the expression more closely
we see that for a fixed $\,\vec a_1\,$ it represents a family of
lines with the common intersection
$$
\left\lbrack\, \frac{4 |a_1\!-a_2|
\sin^{2}b_1\!-1}{2\pi},\,-\,\frac{1}{2\pi} \right\rbrack\,.
$$
Obviously, we can change the number of bound states for fixed
coupling constants by changing the slope of the line (it depends on
$\,b_1,b_2\, $ only) or shifting the line in the
$\,\alpha_1$--direction changing $\,|a_1\!-a_2|\,$.

Similar conclusions can be made for a larger number of perturbations.
Although parametric expressions of the hypersurfaces in $\,\R^N$
dividing regions with different number of bound states are easy to
derive investigating $\, \lim_{z\to 1-} \det \Lambda (\alpha, \vec
a;z)\,$, the formulas are somewhat lengthy and we do not write them
down here. We restrict ourselves to the observation that
$$
P_N(\alpha_1,\ldots,\alpha_N)\,=\,P_{N-1}(\alpha_1,\ldots,\alpha_{N-1})
\,\alpha_N\,+\,Q_{N-1}(\alpha_1,\ldots,\alpha_{N-1})
$$
for some polynomial $\,Q_{N-1}\,$, which means that the dividing
hypersurface for $\,N\,$ perturbations is singular as a function of
$\,\alpha_1,\dots,\alpha_{N-1}\,$ at those points for which
$\,P_{N-1}(\alpha_1,\ldots,\alpha_{N-1})=0\,$, \ie, on the previous
surface. Number of bound states is changed by one each time when the
surface in the coupling--constant space is crossed. It means that
having found the number of bound states for a given setting of
$\,N\!-\!1\,$ centers, we can decide whether adding another center
increases the number of bound states or not: it does if
$\,P_N(\alpha_1,\ldots,\alpha_N)<0\,$, while in the opposite case it
does not.

Another question concerns possible degeneracy of the discrete
spectrum. For the sake of brevity, we write the matrix
(\ref{Lambda_N}) as
$$
\Lambda(\alpha,\vec a;z)_{jm}\,=\,\delta_{jm} (\alpha_j\!- \xi_j(z))
+(1\!-\!\delta_{jm}) g_{jm}(z)
$$
with $\,\xi_j(z):= \xi(\vec a_j,z)\,$ and $\,g_{jm}(z):= -G_0(\vec
a_j, \vec a_m;z)\,$. Since $\,-g_{jm}(z)>0\,$, the {\em maximum
degeneracy} is $\,N\!-\!1\;$; in particular, the discrete spectrum is
always simple for $\,N=2\,$. This is not surprising; we know from
Section~3.2 that the ground state is nondegenerate. On the other
hand, degeneracies may occur. Consider the case $\,N=3\,$ with
$\,\vec a_{1,3}:=\left( \pm a,\, {\pi\over 2}\right)\,$ and $\,\vec
a_2:= (0,b)\,$, and fix an energy $\,z<1\,$. We have $\,g_{12}(z)=
g_{23}(z)\,$ for any $\,b\in(0,\pi)\,$. If $\,b=\,{\pi\over 2}\,$
this value is by (\ref{free kernel}) strictly greater than
$\,g_{13}(z)\;$; on the other hand, $\,\lim_{b\to 0} g_{12}(z)=0\,$,
so there is a $\,b\in \left(0,{\pi\over 2}\right)\,$ for which all
the three $\,g_{jm}(z)\,$ have the same value. Choosing now the
coupling constants $\,\alpha_j\,$ in such a way that
$\,\alpha_j-\xi(\vec a_j;z)= g_{jm}(z)\,,\; j=1,2,3\,$, we find that
$\,z\,$ is an eigenvalue of multiplicity two.

It should be noted, however, that degenerated eigenvalues occur
rather exceptionally. The matrix $\,\Lambda(\alpha,\vec a;z)\,$ has
$\,3N\,$ real parameters, because one of the coordinates $\,a_j\,$
may be chosen arbitrarily, while the number of different $\,2\times
2\,$ submatrices is $\,{N+1\choose 2}\,$. In particular, the number
of the available parameters is always less than the number of
matrices to be annulated for $\,N\ge 6\,$.

Suppose now that we fix $\,N\!-\!1\,$ coupling constants and let the
remainig one run from $\,-\infty\,$ to $\,\infty\,$. The whole
discrete spectrum moves up at that but in a peculiar way; if we do
not hit a degeneracy point by a chance, we observe {\em avoided
crossings} between the energy levels of the $\,(N\!-\!1)$--center
Hamiltonian and the running eigenvalue corresponding to the chosen
perturbation. If all the fixed $\,\alpha_j\,$ represent a strong
coupling, so that they meet the graph of the chosen level at large
negative energies where it is steep, and if the crossings are
narrowly avoided, we reproduce in the present situation the {\em
cascading phenomenon} studied in \cite{GGH}.

To explain the mechanism responsible for this behavior, consider the
case of two perturbations with coupling constants $\,\alpha:=
\alpha_1\,$ and $\,\alpha_2\,$. If one of the perturbations is
absent, the other one has a single eigenvalue, $\,e_1(\alpha)\,$ and
$\,e_2(\alpha)\equiv e_2\,$, respectively. Now we let $\,\alpha\,$
running and look how the eigenvalues influence each other, in
particular, when $\,|\alpha|\,$ is large. One of the solutions to the
condition (\ref{pole_N}) is
   \begin{equation}
\epsilon_2(\alpha)\,=\, e_2+\, {c_2(e_2)\over\alpha_0\!-\alpha}\,+\,
\OO(\alpha^{-2}) \,,
   \end{equation}
where
$$
c_j(z)\,:=\, {g_{12}(z)^2\over \xi'_j(z)}\,>\,0\,, \qquad
\alpha_0\,:=\, \xi_1(e_2)\,-\, {2(g_{12}g'_{12})(e_2) \over
\xi'_2(e_2)}\;;
$$
the first term on the \rhs of the last expression corresponds to the
crossing point, $\,e_1(\alpha)=e_2\,$, and the second one is the
shift due to level interaction. In a similar way, the running
eigenvalue changes to
$$
\epsilon_1(\alpha) \,\approx\, e_1(\alpha)\, +\,
{g_{12}(e_1(\alpha))^2\over \xi'_1(e_1(\alpha))[\xi_2(e_1(\alpha))-
\alpha_2]-  2(g_{12}g'_{12})(e_1(\alpha))}\,.
$$
Using asymptotic properties of $\,\xi_j\,$ together with (\ref{energy
asymptotics}), we find that
   \begin{equation}
\epsilon_1(\alpha)\,\approx\, e_1(\alpha)\,+\, {4\pi\over\alpha}\,
\left(g_{12}(e_1(\alpha))\right)^2 e^{2(\Psi(1)-2\pi\alpha)}
   \end{equation}
as $\,\alpha\to-\infty\;$; recall that $\,g_{12}(e_1(\cdot))\,$
itself is exponentially decaying. On the other hand, there is no
reasonable asymptotics for $\,\alpha\to\infty\,$ because the cross
terms $\,g_{12}(z)\,$ can no longer be regarded as a perturbation
there; in fact, the perturbed eigenvalue may disappear in the
continuum.

Nevertheless, if $\,2g_{12}g'_{12}\,$ remains small in a wide range
of energies, there is an interval $\,(\alpha',\alpha'')\,$ above the
crossing point where the first term in the denominator dominates and
we have
$$
\epsilon_1(\alpha)\,\approx\, e_1(\alpha)\,+\, {c_1(e_1(\alpha))
\over \xi_2(e_1(\alpha))-\tilde\alpha_0}
$$
with $\,\tilde\alpha_0:= \alpha_2+ (2g_{12}g'_{12}/\xi'_1)
(e_1(\alpha))\;$; the correction term is positive. We see therefore
that away of the crossing region the two eigenvalues follow closely
the two branches of the ``decoupled'' spectrum graph. It is also
clear that the smaller are the terms $\,g_{12}\,$ in the above
formulas the closer come the two curves together; since the function
$\,G_0(\vec a_1,\vec a_2;z)\,$ at a fixed energy decrease
exponentially with the distance of the two points, the most profound
cascading effect may be expected when the impurities are far apart.
Moreover, if the perturbations produce a multiple eigenvalue or a
cluster of almost identical simple eigenvalues, the cascading means
that one eigenvalue leaves the cluster and one joins it; this is
illustrated on Figure~7.
   \begin{figure} \label{cascading}
   \caption{The cascading effect for a cluster of eigenvalues}
   \end{figure}

\subsection{Scattering}

The existence and completeness of wave operators follow again from
the Kato--Birman theory. The argument of Section~2.6 can be easily
modified: the generalized eigenfunction (\ref{generalized
eigenfunction}) at a non--integer $\,z>1\,$ has to be replaced by
   \begin{equation} \label{generalized eigenfunction_N}
\psi(\vec x)\,=\, e^{ik_n(z)x} \chi_n(y)\,+\, \sum_{j,k=1}^N
(\Lambda(z)^{-1})_{jk} G_0(\vec x,\vec a_j;z)\, e^{ik_n(z)a_k}
\chi_n(b_k)
   \end{equation}
with the incident wave in the $\,n$--th channel. Consequently, the
reflection and transmission amplitudes from the $\,n$--th to the
$\,m$--th channel are
   \begin{eqnarray} \label{rt_N}
r_{nm}(z) &\!=\!& {i\over\pi}\, \sum_{j,k=1}^N (\Lambda(z)^{-1})_{jk}
{\sin(mb_j) \sin(nb_k)\over k_m(z)}\, e^{i(k_ma_j+k_na_k)}\;,
\nonumber \\ \\
t_{nm}(z) &\!=\!& \delta_{nm}\,+\, {i\over\pi}\, \sum_{j,k=1}^N
(\Lambda(z)^{-1})_{jk} {\sin(mb_j) \sin(nb_k)\over k_m(z)}\,
e^{-i(k_ma_j-k_na_k)}\,. \nonumber
   \end{eqnarray}
The unitarity condition now reads
   \begin{eqnarray} \label{unitarity_N}
\sum_{m=1}^{[\sqrt z]} k_m (t_{nm}\overline t_{sm}+r_{nm}\overline
r_{sm}) &\!=\!& \delta_{ns} k_n\,, \nonumber \\ \\
\sum_{m=1}^{[\sqrt z]} k_m \left(\tilde t_{nm}\overline r_{sm}
+\tilde r_{nm}\overline t_{sm}\right) &\!=\!& 0 \,, \nonumber
   \end{eqnarray}
because $\,S_{nm}\,$ is given again by (\ref{S block}), where the
tilded quantities are obtained by mirror transformation, \ie, by
changing each perturbation longitudinal coordinate $\,a_j\,$ to
$\,-a_j\,$.

The knowledge of the reflection and transmission coefficients makes
it possible to express quantities of a direct physical interest, in
the first place the conductance which is given by the Landauer
formula
   \begin{equation} \label{Landauer}
G(z)\,=\, {2e^2\over h}\, \sum_{n,m=1}^{[\sqrt z\,]}\,
{k_m\over k_n}\, |t_{nm}(z)|^2\,,
   \end{equation}
where $\,t_{nm}(z)\,$ are the coefficients (\ref{rt_N}). An example
is shown on Figure~8;
   \begin{figure} \label{conductance}
   \caption{The conductance plot for a pair of point perturbations.}
   \end{figure}
we see that the perturbations deform the ideal steplike shape. The
most remarkable feature are the sharp resonance peaks and dips which
approach the channel thresholds in the weak--coupling limits. The
corresponding resonance poles can be found as in Section~2.7 but we
shall not go into the details.

The fact that our model is explicitly solvable makes it a useful tool
to study various methods in which the scattering matrix for a system
of perturbations is constructed through a factorization from
single--obstacle objects. In practice such techniques involve a
restriction on the number of evanescent modes taken into account
between subsequent scatterers. A detailed discussion of this
situation in our framework is presented in another paper \cite{ET};
it shows that a particular caution is required in the resonance
regions.

\section{A cylindrical strip in axial magnetic field}

\setcounter{equation}{0}

Before proceeding further, let us show how the above reasoning can be
modified for a different geometry: we shall suppose that the strip is
coiled into the form of a cylinder and placed into a homogeneous
magnetic field parallel to the cylinder axis. Here we consider the
situation when the {\em longitudinal} coordinate is locked cyclically
so the resulting cylinder is of a finite height; the case of an
infinite cylinder which was discussed recently on a heuristic level
in
\cite{KNR} will be treated in the next section.

For a finite--height cylinder the configuration space is compact and
may be mapped onto the rectangle
$$
\Omega\,:=\, [0,2\pi)\,\times\,[0,d]
$$
with the cylinder perimeter equal to $\,2\pi\,$ on a suitable length
scale. The vector potential at the cylinder surface is $\,\vec A=
\left( {1\over 2}B, 0 \right)\,$, so
   \begin{equation} \label{mg flux}
\phi\,:=\, \pi B\,=\,2\pi A
   \end{equation}
is the magnetic flux through the cylinder.

\subsection{The free Hamiltonian}

The free problem can again be solved by separation of variables. It
is convenient to include the magnetic field into boundary conditions:
the unperturbed operator $\,H_0(\phi)\,$ acts then as Laplacian on
$\,\Omega\,$ satisfying the condition (\ref{strip boundary}) for all
$\,x\in[0,2\pi)\,$ together with
   \begin{equation} \label{mg bc}
\psi(0+,y)\,=\,e^{-i\phi}\psi(2\pi-,y)\,, \qquad
{\partial\psi\over\partial x}(0+,y)\,=\,
e^{-i\phi}\,{\partial\psi\over\partial x} (2\pi-,y)
   \end{equation}
for all $\,y\in [0,d]\,$. Its spectrum is purely discrete (\ie,
finitely degenerated) and consists of the eigenvalues
   \begin{equation} \label{free eigenvalues}
\epsilon_{mn}\,:=\, (m+A)^2+(\nu n)^2\,, \quad m\in\Z\,,\; n=1,2,\dots\;;
\qquad \nu\,:=\,{\pi\over d}\;.
   \end{equation}
In particular, if $\,\nu^2\,$ is irrational the spectrum is simple
for $\,2A\not\in\Z\,$ and twice degenerated otherwise (with the
exception of the ground state for $\,A\in\Z\,$). The corresponding
eigenfunctions are of the form $\,\eta_m\otimes \chi_n\,$, where
$\,\chi_n\,$ are the transverse--mode functions used above and
$$
\eta_m(x)\,:=\, {1\over \sqrt{2\pi}}\, e^{i(m+A)x}\,, \quad m\in\Z\,.
$$
It is straightforward to write the unperturbed resolvent kernel
   \begin{equation} \label{free mg resolvent 1}
G_0(\vec x_1,\vec x_2;\phi,z)\,=\, {1\over \pi d}\,
\sum_{n=1}^{\infty}\, \sum_{m=-\infty}^{\infty}\,
{e^{i(m+A)(x_1-x_2)} \sin(\nu ny_1)\sin(\nu ny_2) \over \epsilon_{mn}-z} \,.
   \end{equation}
The \rhs has to be handled with caution because the series is not
absolutely convergent. Fortunately, one can get rid of the double
summation by evaluating the inner series: we have
$$
\sum_{m=-\infty}^{\infty}\,{e^{imx}\over (m+A)^2+\gamma}  \,=\,
{\pi\over 2i\sqrt\gamma}\, \left( {e^{i(A-i\sqrt\gamma)(\pi-x)} \over
\sin(\pi(A-i\sqrt\gamma))}\,-\, {e^{i(A+i\sqrt\gamma)(\pi-x)} \over
\sin(\pi(A+i\sqrt\gamma))} \right)
$$
for $\,x\in[0,2\pi)\;$ \cite[5.4.3.4]{PBM}, while for
$\,x\in(-2\pi,0)\,$ one has to replace $\,\pi-x\,$ by $\,-\pi-x\;$;
this yields
   \begin{eqnarray} \label{free mg resolvent 2}
G_0(\vec x_1,\vec x_2) &\!=\!&
\sum_{n=1}^{\infty}\, {\sinh((2\pi\!-\!|x_1\!-\!x_2|)
\sqrt{\nu^2n^2\!-\!z}) +e^{2i\eta\pi A} \sinh(|x_1\!-\!x_2|
\sqrt{\nu^2n^2\!-\!z}) \over \cosh(2\pi\sqrt{\nu^2n^2\!-\!z})
-\cos(2\pi A)} \nonumber \\ \\ &&
\times\:{\sin(\nu ny_1)\sin(\nu ny_2) \over
d\sqrt{\nu^2n^2\!-\!z}}\,, \nonumber
   \end{eqnarray}
where $\,\eta:=\sgn(x_1\!-x_2)\,$.

\subsection{The perturbed resolvent}

As before we add now to the operator $\,H_0(\phi)\,$ a finite number
of point perturbations with the coupling constants $\,\alpha_1,
\dots,\alpha_N\,$ placed at $\,\vec a_1,\dots, \vec a_N\in\Omega\;$;
the corresponding Hamiltonian $\,H(\alpha,\vec a,\phi)\,$ is again
determined by the boundary conditions (\ref{bc_N}). To find its
resolvent we repeat the above argument writing a general vector from
the domain of $\,H(\alpha,\vec a,\phi)\,$ in the form
(\ref{decomposition_N}). In view of (\ref{free mg resolvent 2}), we
have
$$
\lim_{|x-a_j|\to 0}\, {G_0(\vec x,\vec a_j)\over \ln|x-a_j|}
\,=\, {1\over\pi}\, \lim_{y\to b_j}\, {1\over \ln|y-b_j|}\,
\sum_{n=1}^{\infty}\, {\sin(\nu ny_1)\sin(\nu ny_2) \over n}\,
\coth(\pi\nu n)\,.
$$
Since $\,\coth(\pi\nu n)\,$ tends to one exponentially fast as
$\,n\to\infty\,$, it may be neglected; Using further the identity
$$
\sum_{n=1}^{\infty}\, {\sin(\nu ny_1)\sin(\nu ny_2) \over n}\,=\,
{1\over 4}\, \ln\left(\left( {\sin\left(\nu{y+b_j\over 2}\right)
\over \sin\left(\nu{y-b_j\over 2}\right)} \right)^2 \right)
$$
\cite[5.4.15.1]{PBM} we find that $\,L_0(\psi,\vec a_j)\,$ is given
by the same expression as in Section~3.1. In this way, the sought
resolvent kernel is given by a formula analogous to (\ref{Krein_N})
with
   \begin{equation} \label{lambda mg}
\lambda(\alpha,\vec a,\phi;z)\,=\, \Lambda(\alpha,\vec
a,\phi;z)^{-1}\,,
   \end{equation}
where $\,\Lambda_{jr}= -G_0(\vec a_j,\vec a_r)\,$ for $\,j\ne r\,$,
while the diagonal elements are given by
$$
\Lambda_{jj}\,=\, \alpha_j\,-\, \lim_{|x-a_j|\to 0}\, \left(G_0(\vec
x,\vec a_j) +\,{1\over 2\pi}\, \ln|x-a_j| \right)\,.
$$
The limit is easily evaluated using (\ref{free mg resolvent 2}); we
arrive at the expressions
   \begin{eqnarray} \label{Lambda mg}
\Lambda_{jj} &\!=\!& \alpha_j\,-\, \sum_{n=1}^{\infty}\, \left(
{\sinh(2\pi\sqrt{\nu^2n^2-z}) \over \cosh(2\pi\sqrt{\nu^2n^2-z})
-\cos\phi}\: {\sin^2(\nu nb_j)\over d\sqrt{\nu^2n^2-z}}\,-\, {1\over
2\pi n} \right)\,, \nonumber \\ \\
\Lambda_{jr} &\!=\!& -G_0(\vec a_j,\vec a_r)\,, \qquad j\ne r\;;
\nonumber
   \end{eqnarray}
notice that the matrix $\,\Lambda(z)\,$ does not change if the flux
$\phi\,$ is modified by an integer multiple of $\,2\pi\,$.

\subsection{Spectral properties}

Since the point interactions represent a finite--rank perturbation to
the resolvent, $\,\sigma_{ess}(H(\alpha,\vec a,\phi))=
\sigma_{ess}(H_0(\phi))= \emptyset\,$, \ie, the spectrum of
$\,H(\alpha,\vec a,\phi)\,$ is again purely discrete.

Moreover, it is disjoint with $\,\EE:=\{\,\epsilon_{mn}:\: m\in\Z,\:
n=1,2,\dots\,\}\,$ provided there is at least one point perturbation
present. If some $\,\epsilon_{mn}\,$ would be an eigenvalue of
$\,H(\alpha,\vec a,\phi)\,$, the corresponding eigenfunction had to
satisfy the corresponding Helmholz equation everywhere outside the
set $\,\{\vec a_1,\dots,\vec a_N\}\,$ being therefore a {\em finite}
linear combination $\,\psi= \sum_{m,n} c_{mn} \eta_m\otimes
\chi_n\,$. Such a function, however, is bounded in $\,\Omega\,$ and
cannot have a logarithmic singularity, \ie, $\,L_0(\psi,\vec a_j)
=0\,$ for any $\,j\,$.

A real number $\,z\not\in \EE\,$ therefore belongs to the resolvent
set of $\,H(\alpha,\vec a,\phi)\,$, so we are allowed to use the
argument of Section~3.3. The eigenvalues are given by the condition
   \begin{equation} \label{pole mg}
\det \Lambda(\alpha,\vec a,\phi;z)\,=\,0\,,
   \end{equation}
under which the linear system
$$
\sum_{r=1}^N\, \Lambda(z)_{jr} d_r\,=\,0
$$
has a solution (or solutions); the corresponding (non--normalized)
eigenfunctions are then
   \begin{equation} \label{eigenfunction mg}
\varphi(\vec x)\,=\, \sum_{r=1}^N\, d_j G_0(\vec x,\vec
a_j;\phi,z)\,.
   \end{equation}

\section{Point perturbations on an infinite cylinder}

\setcounter{equation}{0}

If the cylinder heigth $\,d\,$ of the previous section goes to
infinity, the configuration space becomes again noncompact; one can
map it on the strip
$$
\Omega\,=\, [0,2\pi)\times\R\,.
$$
The results of Section~\ref{finite number} modify easily to the
present situation.

\subsection{The resolvent and scattering amplitudes}

The free Hamiltonian is defined again by the cyclic boundary
conditions (\ref{mg bc}), so the ``transverse eigenfunctions'' are
$\,\eta_m:\: \eta_m(x)=\, (2\pi)^{-1/2} e^{i(m+A)x}\,$ corresponding
to the eigenvalues $\,\epsilon_m:=\,(m+A)^2\,,\; m\in\Z\,$. The free
resolvent kernel is therefore
   \begin{equation} \label{free infinite mg resolvent}
G_0(\vec x_1,\vec x_2;\phi,z)\,=\, {i\over 4\pi}\,
\sum_{m\in\Z}\, {e^{i(m+A)(x_1-x_2)} e^{ik_m(z)|y_1-y_2|} \over
k_m(z)}\,,
   \end{equation}
where
   \begin{equation} \label{mg momentum}
k_m(z)\,:=\, \sqrt{z-(m+A)^2}\,.
   \end{equation}
Introducing now $\,N\,$ point perturbations as in Section~\ref{finite
number}, we find that the full resolvent kernel is given by the
relations (\ref{Krein_N}) and (\ref{lambda_N}) with
   \begin{eqnarray} \label{Lambda infinite mg}
\Lambda_{jj} &\!=\!& \alpha_j\,-\, {i\over 4\pi}\, \left\lbrack\,
{1\over k_0(z)}\,+\, \sum_{m\ne 0}\, \left( {1\over k_m(z)}+
{i\over|m|}  \right) \right\rbrack\,, \nonumber \\ \\
\Lambda_{jk} &\!=\!& -G_0(\vec a_j,\vec a_k)\,, \qquad j\ne k\;;
\nonumber
   \end{eqnarray}
It is convenient for computational purposes to rewrite the first of
these relations in the form
   \begin{equation} \label{Lambda infinite mg2}
\Lambda_{jj} \,=\, \tilde\alpha_j\,-\, {i\over 4\pi}\, \left\lbrack\,
{1\over k_0(z)}\,+\, \sum_{m\ne 0}\, \left( {1\over k_m(z)}+
{i\over|m\!+\!A|}  \right) \right\rbrack\,,
   \end{equation}
where $\,\alpha_j\,$ are renormalized coupling constants,
$$
\tilde\alpha_j\,=\, \alpha_j\,+\, {1\over 4\pi}\, \sum_{m\ne 0}\,
\left( {1\over |A|}\,-\,{1\over|m\!+\!A|}  \right)\;;
$$
we suppose, of course, that $\,A\not\in\Z\,$.

From here one can find the discrete spectrum of $\,H(\alpha,\vec
a,\phi)\,$ by virtue of the eqs. (\ref{pole mg}) and
(\ref{eigenfunction mg}). Similarly, the reflection and transmission
amplitudes are
   \begin{eqnarray} \label{rt_N mg}
r_{nm}(z) &\!=\!& {i\over 4\pi}\, \sum_{j,k=1}^N
(\Lambda(z)^{-1})_{jk} {e^{-i(m+A)b_j} e^{i(n+A)b_k} \over k_m(z)}\,
e^{i(k_ma_j+k_na_k)}\;,
\nonumber \\ \\
t_{nm}(z) &\!=\!& \delta_{nm}\,+\, {i\over 4\pi}\, \sum_{j,k=1}^N
(\Lambda(z)^{-1})_{jk} {e^{-i(m+A)b_j} e^{i(n+A)b_k} \over k_m(z)}\,
e^{-i(k_ma_j-k_na_k)}\;; \nonumber
   \end{eqnarray}
the elementary block $\,S_{nm}\,$ of the S--matrix is given again by
(\ref{S block}), where the tilded quantities are obtained by mirror
transformation. The scattering is, of course, a one--dimensional
problem with an additional degree of freedom corresponding to the
transverse channels --- it imposes no restriction on the size of the
cylinder contrary to the claim made in \cite{KNR}.

\section{An infinite strip in external field}

\setcounter{equation}{0}

Let us return now to the infinite strip and assume that, in addition
to point perturbations, the particle interacts with a potential. Such
a situation can occur, for instance, if the strip is cut not from a
plane but from a curved surface (see Fig.~9)  
   \begin{figure} \label{gendarme acouchee}
   \caption{A strip potential due to an external field.}
   \end{figure}
and exposed to a homogeneous electric field. We shall suppose that
the potential depends only on the longitudinal variable; otherwise
the problem would become much more complicated.

\subsection{The unperturbed Hamiltonian}

In the described situation the Hamiltonian free of point
perturbations is therefore
   \begin{equation} \label{free Hamiltonian e}
H_V\,:=\, -\partial_x^2-\partial_y^2+V(x)
   \end{equation}
with the usual Dirichlet condition (\ref{strip boundary}) at the
strip boundary. Our assumptions about the potential are for the
moment rather weak: we require only that $\,V\in L^1_{loc}(\R)\,$ and
the operator
$$
h_V\,:=\, -\partial_x^2+V(x)
$$
with the natural domain is \esa$\,$ Then $\,H_V\,$ understood as the
closure of $\,h_V\otimes(-\partial_y^2)\,$ is self--adjoint and its
spectrum is
   \begin{equation} \label{free spectrum e}
\sigma(H_V)\,=\, \{\, \lambda+n^2\,:\; \lambda\in\sigma(h_V)\,,\:
n=1,2,\dots\, \}\,.
   \end{equation}
In particular, the resolvent set $\,\rho(H_V)\supset (-\infty,
V_-+1)\,$, where $\,V_-:= \inf_{x\in\R} V(x)\,$. Of course, the
character of the spectrum depends strongly on that of
$\,\sigma(H_V)\,$. If, for example, the potential vanishes at
infinity, $\,\lim_{|x|\to\infty} V(x)=0\,$, then $\,h_V\,$ is below
bounded with $\,\sigma_{ess}(h_V)= [0,\infty)\,$ by \cite[Secs.X.2
and XIII.4]{RS}, so
$$
\sigma_{ess}(H_V)= [1,\infty)
$$
as in the absence of the potential. In a similar way, if the
potential grows at large distances (at least in the mean) then the
spectrum of $\,H_V\,$ is pure point \etc

By assumption, $\,h_V\,$ is limit point at both $\,\pm\infty\,$, so
it follows from the Weyl alternative that to any $\,z=k^2\in
\C\setminus\R\,$ there are functions $\,u\equiv u(\cdot,k)\,$ and
$\,v\equiv v(\cdot,k)\,$, unique up to a multiplicative constant,
which solve the equation $\,f''+(k^2-v)f=0\,$ being square integrable
at $\,\mp\infty\,$, respectively. They are pointwise continuous with
respect to $\,k\;$; we use the same symbols for their extensions to
the positive real axis in the $\,k$--plane. We denote $\,k_n(z):=
\sqrt{z-n^2}\,$ as before and set
$$
u_n\,:=\,u(\cdot,k_n(z))\,, \qquad v_n\,:=\,v(\cdot,k_n(z))\,.
$$
The free resolvent kernel is then
   \begin{equation} \label{free kernel e}
G_V(\vec x_1,\vec x_2;z)\,=\, -\,{2\over\pi}\, \sum_{n=1}^{\infty}\,
{u_n(x_<)v_n(x_>) \over W(u_n,v_n)}\, \sin(ny_1)\sin(y_2)\,,
   \end{equation}
where $\,x_<,\,x_>\,$ is the smaller and larger of $\,x_1,\,x_2\,$,
respectively.

\subsection{Point perturbations}

The Hamiltonian $\,H(\alpha,\vec a,V)\,$ with point perturbations is
defined again through the boundary conditions (\ref{bc_N}). To find
its resolvent, we need the logarithmic singularity of the kernel
(\ref{free kernel e}). If the potential is smooth the latter is known
to be the same as in the absence of the potential \cite{Ti}. In fact
the potential need not be regular; nevertheless an additional
restriction is required because the property under consideration is
local, and therefore could be affected by a local singularity. We
shall assume that the potential is {\em locally bounded}, \ie,
bounded in any compact interval.

To find the singularity, we notice that what matters here is the
behavior of the series at large $\,n\;$; however, these terms
correspond to solutions which are deep in the classically forbidden
region (\ie, with the imaginary part of the {\em channel} momentum
large positive) and therefore of the WKB--form,
$$
u_n(x)= \kappa_n(x)^{-1/2}\,e^{\int_0^{\infty}
\kappa_n(x')dx'}\, \err_{1}(\kappa_n)\,, \;\,
v_n(x)= \kappa_n(x)^{-1/2}\,e^{-\int_0^{\infty}
\kappa_n(x')dx'}\, \err_{1}(\kappa_n)\,,
$$
where $\,\kappa_n(x):= \sqrt{n^2+V(x)-z}\,$ and
$\,\err_{\alpha}(\kappa)\,$ stands as a shorthand for
$\,1+\OO(\kappa^{\alpha})\,$, so
$$
{u_n(x)v_n(x) \over W(u_n,v_n)}\,=\, -\,{1\over 2\kappa_n(x)}\,
\err_1(\kappa_n) \,=\, -\,{1\over 2n}\,\err_1(n)
$$
due to the local boundedness of the potential. Hence we obtain
$$
\lim_{|\vec x-\vec a_j|\to 0}\: {G_V(\vec x,\vec a_j;z)\over \ln(\vec
x-\vec a_j|}\,=\, -\,{1\over 2\pi}\,,
$$
and the argument of Section~3.1 can be easily modified giving
   \begin{equation} \label{Krein e}
(H(\alpha,\vec a,V)\!-\!z)^{-1}(\vec x_1,\vec x_2)\,=\, G_V(\vec
x_1,\vec x_2;z) +\sum_{j,k=1}^N (\Lambda(z)^{-1})_{jk} G_V(\vec
x_1,\vec a_j;z) G_V(\vec a_k,\vec x_2;z)\,,
   \end{equation}
where $\,\Lambda(\alpha,\vec a,V;z)\,$ is the following $\,N\times
N\,$ matrix:
   \begin{eqnarray} \label{Lambda e}
\Lambda_{jr} &\!=\!& -\,G_V(\vec a_j,\vec a_r;z)\,, \qquad j\ne r\,,
\nonumber \\ \\
\Lambda_{jj} &\!=\!& \alpha_j\,+\, {2\over\pi}\, \sum_{n=1}^{\infty}
\left( {u_n(a_j) v_n(a_j)\over W(u_n,v_n)}\, \sin^2(nb_j) \,+\,
{1\over 4n} \right)\,. \nonumber
   \end{eqnarray}
It becomes singular if the condition
   \begin{equation} \label{pole e}
\det \Lambda(\alpha,\vec a,V;z)\,=\,0
   \end{equation}
is valid; the corresponding $\,z\,$ are eigenvalues of
$\,H(\alpha,\vec a,V)\,$. If $\,z\not\in \rho(H_V)\,$, the argument
of Section~3.2 may be used: the non--normalized eigenfunctions are of
the form
   \begin{equation} \label{eigenfunction e}
\varphi(x)\,=\, \sum_{j=1}^N d_j G_V(\vec x,\vec a_j;z)\,,
   \end{equation}
where $\,d\in\C^N\,$ is the respective solution to $\,\sum_{k=1}^N
\Lambda(z)_{jk} d_k=0\,$.

\subsection{Scattering}

One has to assume that the underlying scattering problem for the
potential $\,V\,$ makes sense, the wave operators exist and are
complete; since $\,V\,$ is locally bounded by assumption, this is
ensured, for instance, if
$$
|V(x)|\,\le\, C(1+|x|)^{-1-\eps}\,.
$$
for some positive $\,C,\:\eps\;$ \cite[Sec.XI.4]{RS}. Repeating the
limiting--absorption argument of Secs.2.6 and 3.4, we get the
generalized eigenfunction
$$
\psi(\vec x)\,=\, e^{ik_n(z)x}\chi_n(y)\,+\, \sum_{j,k=1}^N
(\Lambda(z)^{-1})_{jk} G_V(\vec x,\vec a_j;z)\, e^{ik_n(z)a_k}
\chi_n(b_k)\,.
$$
We substitute for $\,G_V\,$ from (\ref{free kernel e}) and consider
the asymptotic behavior. To this end, we need the asymptotics of the
elementary solutions which is given by
   \begin{equation} \label{x scattering}
u_n(x)\,\approx\, \left\{\; \begin{array}{lll} e^{-ik_n(z)x} \quad &
\dots \quad & x\to-\infty \\
\overline\tau_n\, e^{-ik_n(z)x}+ \overline\rho_n\,e^{ik_n(z)x} \quad &
\dots \quad & x\to\infty \end{array} \right.
   \end{equation}
and an analogous formula with the interchange of signs for
$\,v_n(x)\;$; here $\,\tau_n:= \tau(k_n(z))\,$ and $\,\rho_n:=
\rho(k_n(z))\,$ are the standard (left--to--right) transmission and
reflection amplitudes for the potential $\,V\,$ at the momentum
$\,k_n(z)\,$. This yields
   \begin{eqnarray} \label{rt e}
r_{nm}(z) &\!=\!& {i\over\pi}\, \sum_{j,k=1}^N (\Lambda(z)^{-1})_{jk}
{\sin(mb_j) \sin(nb_k)\over k_m(z) \overline\tau(k_m(z))}\,
u_m(a_j)\, e^{ik_n(z)a_k}\;,
\nonumber \\ \\
t_{nm}(z) &\!=\!& \delta_{nm}\,+\, {i\over\pi}\, \sum_{j,k=1}^N
(\Lambda(z)^{-1})_{jk} {\sin(mb_j) \sin(nb_k)\over k_m(z)
\overline\tau(k_m(z))}\, v_m(a_j)\, e^{ik_n(z)a_k}\;; \nonumber
   \end{eqnarray}
the unitarity condition (\ref{unitarity_N}) remains the same.

\section{Periodic perturbations}

\setcounter{equation}{0}

An extension of the results discussed above to an infinite number of
perturbations is straightforward for the algebraic part; the matrices
$\,\Lambda\,$ become infinite--dimensional, \ie, operators on the
Hilbert space $\,\ell^2$. However, since they are generally
unbounded, the corresponding analysis is much harder. The problem
should be treated along the same lines as in \cite[Sec.III.1.1]{AGHH}
but we are not going to do that; instead we concentrate on an
important particular case of a strip in which the point perturbations
are periodically arranged.

\subsection{The Floquet--Bloch analysis}

Suppose that the set of point perturbations $\,\{\alpha,\vec
a\}\equiv \{[\alpha_j,\vec a_j]:\; j=1,2,\dots\,\}\,$ in $\,\Omega\,$
is countably infinite and has a periodic pattern with a period
$\,\ell>0\,$. As usually we suppose that the width of the strip is
$\,d=\pi\;$; the general case can be obtained by a simultaneous
rescaling of the coupling constants.

The existence of the corresponding operator $\,H(\alpha,\vec a)\,$
can be established as in \cite[Sec.III.1.1]{AGHH}. Following the
standard procedure named Floquet by mathematicians and Bloch by
physicists, we can find a unitary operator $\,U:\, L^2(\Omega)\to
L^2(\BB, (\ell/2\pi)d\theta; L^2(\hat\Omega))\,$, where
   \begin{equation} \label{WS cell}
\hat\Omega\,:=\, [0,\ell]\times [0,\pi]\,, \qquad \BB\,:=\,
\left\lbrack\, -{\pi\over\ell},\,{\pi\over\ell}\right) \times
[0,\pi] \;;
   \end{equation}
the $\,x$--projections of these rectangles are the Wigner--Seitz cell
of the underlying one--dimensional lattice, and the corresponding
Brillouin zone, respectively --- \cf\cite[Sec.III.1.3]{AGHH}. By
means of $\,U\,$, the operator $\,H(\alpha,\vec a)\,$ is unitarily
equivalent to
   \begin{equation} \label{FB decomposition}
U\,H(\alpha,\vec a)\,U^{-1}\,=\, {\ell\over 2\pi}\,
\int_{|\theta\ell|\le\pi}^{\oplus} H(\alpha,\vec
a;\theta)\,d\theta\,,
   \end{equation}
where $\,H(\alpha,\vec a;\theta)\,$ is the point--interaction
Hamiltonian on $\,L^2(\hat\Omega)\,$, \ie, the Laplacian satisfying
the boundary conditions (\ref{bc}) for $\,x\in[0,\ell]\,$,
   \begin{equation} \label{Bloch bc}
\psi(\ell-,y)\,=\,e^{i\theta\ell} \psi(0+,y)\,, \qquad
{\partial\psi\over\partial x}(\ell-,y)\,=\,e^{i\theta\ell}\;
{\partial\psi\over\partial x}(0+,y)
   \end{equation}
for $\,y\in[0,\pi]\,$ and (\ref{bc_N}); with an abuse of notation we
shall use the symbol $\,(\alpha,\vec a)\,$ for the subset of the point
interactions in $\,\hat\Omega\,$ and suppose that their number is
$\,N\,$.

The operators $\,H(\alpha,\vec a;\theta)\,$ are easy to investigate
because up to a change in notation and a simple rescaling they
coincide with the magnetic cylinder Hamiltonians of Section~4. In
particular, the ``free'' eigenvalues
   \begin{equation} \label{FB free eigenvalues}
\epsilon_{mn}(\theta)\,:=\, \left({2\pi
m\over\ell}\,+\theta\right)^2 +n^2\,, \quad m\in\Z\,,\; n=1,2,\dots\,,
   \end{equation}
correspond to the eigenfunctions $\,\eta^{\theta}_m\otimes\chi_n\,$,
where $\,\chi_n\,$ are the usual transverse solutions and
$$
\eta^{\theta}_m(x)\,:=\, {1\over \sqrt{\ell}}\, e^{i(2\pi
m+\theta\ell)x/\ell }\,, \quad m\in\Z\,.
$$
Moreover, the free resolvent kernel is
   \begin{eqnarray} \label{FB free resolvent}
G_0(\vec x_1,\vec x_2;\theta;z) &\!=\!&
\sum_{n=1}^{\infty}\, {\sinh((\ell\!-\!|x_1\!-\!x_2|)
\sqrt{n^2\!-\!z}) +e^{2i\eta\theta\ell} \sinh(|x_1\!-\!x_2|
\sqrt{n^2\!-\!z}) \over \cosh(\ell\sqrt{n^2\!-\!z})
-\cos(\theta\ell)} \nonumber \\ \\ &&
\times\:{\sin(ny_1)\sin(ny_2) \over
\pi\sqrt{n^2\!-\!z}}\,, \nonumber
   \end{eqnarray}
where $\,\eta:=\sgn(x_1\!-x_2)\,$, and the full kernel is expressed
by a formula analogous to (\ref{Krein_N}) with
   \begin{equation} \label{FB lambda}
\lambda(\alpha,\vec a,\theta;z)\,=\, \Lambda(\alpha,\vec
a,\theta;z)^{-1}\,,
   \end{equation}
where $\,\Lambda_{jr}= -G_0(\vec a_j,\vec a_r)\,$ for $\,j\ne r\,$,
while the diagonal elements are given by
   \begin{equation} \label{FB Lambda}
\Lambda_{jj} \,=\, \alpha_j\,-\,{1\over \pi}\, \sum_{n=1}^{\infty}\,
\left( {\sinh(\ell\sqrt{n^2\!-\!z}) \over \cosh(\ell\sqrt{n^2\!-\!z})
-\cos\theta\ell}\: {\sin^2 nb_j\over \sqrt{n^2\!-\!z}}\,-\, {1\over
2n} \right)\,.
   \end{equation}
We may also write the last formula as
$\,\Lambda_{jj}(\alpha,\vec a,\theta;z) =\alpha_j- \xi(\vec
a_j,\theta;z)\,$, where the function $\,\xi\,$ is more explicitly
given by
   \begin{eqnarray} \label{FB xi}
\xi(\vec a_j,\theta;z) &\!=\!& {1\over \pi}\, \sum_{n=1}^{n[z]}\,
\left( {\sin(\ell\sqrt{z\!-\!n^2}) \over \cos(\ell\sqrt{z\!-\!n^2})
-\cos\theta\ell}\: {\sin^2 nb_j\over \sqrt{z\!-\!n^2}}\,-\, {1\over
2n} \right) \nonumber \\ \\
&\!+\!& {1\over \pi}\, \sum_{n=n[z]+1}^{\infty}\,
\left( {\sinh(\ell\sqrt{n^2\!-\!z}) \over \cosh(\ell\sqrt{n^2\!-\!z})
-\cos\theta\ell}\: {\sin^2 nb_j\over \sqrt{n^2\!-\!z}}\,-\, {1\over
2n} \right)\;; \nonumber
   \end{eqnarray}
we have denoted here $\,n[z]:=\min\{0,[\sqrt{z}]\}\,$. It is defined
everywhere except at
   \begin{equation} \label{FB singularities}
\EE(\vec a,\theta)\,:=\, \{\,
\epsilon_{mn}(\theta)\in\EE(\theta)\,:\; \sin nb\ne 0\;\}\,,
   \end{equation}
where $\,\EE(\theta)\,$ is the eigenvalue set (\ref{FB free
eigenvalues}); it is also easy to see that the function $\,\xi\,$ is
monotonously increasing between any pair of neighboring
singularities.

To find the spectrum of the original operator, one has to solve first
the spectral problem for $\,H(\alpha,\vec a;\theta)\,$ in analogy
with the magnetic cylinder case: the eigenvalues are given by
   \begin{equation} \label{FB pole}
\det \Lambda(\alpha,\vec a,\theta;z)\,=\,0\,,
   \end{equation}
and the corresponding eigenfunctions are given by a relation similar
to (\ref{eigenfunction mg}). Next one has to overlay the obtained
eigenvalue sets for $\,\theta\,$ running through the
$\,x$--projection of the Brillouin zone to get the spectrum of
$\,H(\alpha,\vec a)\,$. The main question is, of course, whether we
cover a single interval in this way as in the free case, or whether
the interaction will open some gaps.

\subsection{A single array of perturbations}

Since it is not easy to draw general conclusions about the form of
the spectrum, let us restrict to the simplest case where $\,N=1\,$,
\ie, the strip $\,\Omega\,$ contains a single periodic array of point
perturbation of a strength $\,\alpha\,$ and spacing $\,\ell\,$.
The condition (\ref{FB pole}) then simplifies to
   \begin{equation} \label{FB pole_1}
 \xi(\vec a,\theta;z)\,=\,\alpha\,.
   \end{equation}
We have seen that the function on the left hand side is monotonously
increasing between its singularities, \ie, the points of $\,\EE(\vec
a,\theta)\,$. This means that for fixed $\,\alpha,\,\theta\,$ there
is a sequence $\,\{\epsilon_r(\alpha,\vec a,\theta)\}_{r=0}
^{\infty}\,$ arranged in the ascending order; each of them depends,
in fact, only on the $\,y$--component $\,b\,$ of the vector $\,\vec
a\,$. The lowest one satisfies
$$
\epsilon_0(\alpha,\vec a,\theta)\,<\,1+\theta^2\,,
$$
and between each two neighboring points of $\,\EE(\vec a,\theta)\,$
there is just one of the other eigenvalues. It is also clear that any
of $\,\epsilon_r(\alpha,\vec a,\theta)\,$ is continuous with respect
to the parameters and $\,\epsilon_r(\cdot,\vec a,\theta)\,$ is
increasing for fixed $\,b\,$ and $\,\theta\,$.

On the other hand, the eigenvalues need not be monotonous with
respect to $\,\theta\,$. The implicit--function theorem tells us,
however, that
$$
{\partial\epsilon_r(\alpha,\vec a,\theta)\over \partial\theta}\,=\,
-\, \left.{\partial\xi(\vec a,\theta;z)\over \partial\theta}\, \left(
{\partial\xi(\vec a,\theta;z)\over \partial z} \right)^{-1}\,
\right|_{(\epsilon_r,\theta)}
$$
whenever the denominator is nonzero. Leaving aside the thresholds,
$\,z=n^2$, and the points of $\,\EE(\theta)\,$, a straightforward
differentiation shows that $\,\xi(\vec a,\cdot;\cdot)\,$ is analytic
in both variables. Since the function $\,\xi\,$ is not identically
zero, the derivative $\,\partial\epsilon_r(\alpha,\vec a,\theta)/
\partial\theta\,$ may be zero at some points but never in an
interval. This means that the spectrum of $\,H(\alpha,\vec a)\,$ is
{\em absolutely continuous} --- \cf\cite[Sec.XIII.16]{RS}.

Let us turn now to the question about the number of gaps. Below
$\,z=1\,$ the spectrum may be estimated by means of extrema of the
function $\,\xi\,$: we have
$$
\xi_+(\vec a,z)\,:=\, \max_{|2\theta\ell|\le\pi} \xi(\vec a,\theta;z)
\,=\, {1\over \pi}\, \sum_{n=1}^{\infty}\,\left( {\sin^2 nb \over
\sqrt{n^2\!-\!z}}\, \coth\left({\ell\over 2}\sqrt{n^2\!-\!z} \right)
\,-\, {1\over 2n} \right)\,,
$$
and a similar formula for the minimum, $\,\xi_-(\vec a,z)\,$, with
$\,\coth\,$ replaced by $\,\tanh\,$. Both functions are increasing
and have the same logarithmic asymptotics as $\,\xi(\vec a,z)\,$ of
Section~2.5 as $\,z\to -\infty\,$. On the other hand, $\,\xi_+(\vec
a,\cdot)\,$ diverges as $\,z\to 1-\,$, while $\,\xi_-(\vec
a,\cdot)\,$ has a finite limit. This shows that a gap exists provided
   \begin{eqnarray} \label{one gap}
\alpha\,<\, \xi_-(\vec a,1-) &\!=\!& {\ell\over 2\pi}\, \sin^2 b
-\,{1\over 2\pi} \nonumber \\ \\
&\!+\!& {1\over \pi}\,\sum_{n=2}^{\infty}\,\left( {\sin^2 nb \over
\sqrt{n^2\!-\!1}}\, \tanh\left({\ell\over 2}\sqrt{n^2\!-\!1} \right)
\,-\, {1\over 2n} \right)\,. \nonumber
   \end{eqnarray}
The condition is satisfied for a strong enough coupling, or
alternatively, for any fixed $\,\alpha\,$ and the spacing $\,\ell\,$
large enough.

This conclusion calls to mind the spectrum of a straight polymer in a
plane described in \cite[Sec.III.4]{AGHH} which has also one gap if
the coupling is stronger than some critical value. However, for a
``coated polymer'', \ie, our strip with an array of perturbations a
much stronger result is valid; we are going to show that under a
suitable choice of parameters it can have {\em any finite number} of
gaps.

To this end, we consider $\,z\in(2,3)\,$ and $\,\ell\gg 1\,$, and
rewrite the right hand side of the relation (\ref{FB xi}) as
$$
\xi(\vec a,\theta;z)\,=\, \xi_0(\vec a,\theta;z)+ \eta(\vec
a,\theta;z)\,,
$$
where
$$
\xi_0(\vec a,\theta;z)\,:=\, {\sin(\ell\sqrt{z\!-\!1}) \over
\cos(\ell\sqrt{z\!-\!1}) -\cos\theta\ell}\; {\sin^2 b\over
\pi\sqrt{z\!-\!1}}
$$
and $\,\eta(\vec a,\theta;z)\,$ is the rest. In the same way as
above, one can show that $\,\eta(\vec a,\theta;\cdot)\,$ is
monotonously increasing with a bounded derivative everywhere below
the second threshold, in particular, in the chosen interval of
energies. Moreover,
$$
\eta(\vec a,\theta;z)\,\le\, \eta_+(z)\,:=\,
-\,{1\over 2\pi} \,+\, {1\over \pi}\,\sum_{n=2}^{\infty}\,\left(
{\sin^2 nb \over  \sqrt{n^2\!-\!z}}\, \coth\left({\ell\over
2}\sqrt{n^2\!-\!z} \right)\,-\, {1\over 2n} \right)\,,
$$
and the minimum, $\,\eta_-(z)\,$, is obtained when $\,\coth\,$ is
replaced by $\,\tanh\,$. These estimates become close if $\,\ell\,$
is large; using the inequality $\,\coth u-\tanh u< 5\, e^{-2u}\,$
for $\,2u\ge 1\,$,
we find
$$
\eta_+(z)-\eta_-(z)\,<\, {5\over\pi}\, \sum_{n=2}^{\infty}\,
{\sin^2 nb \over  \sqrt{n^2\!-\!z}}\, e^{-\ell\sqrt{n^2\!-\!z}} \,<\,
{5\over\pi}\, \sum_{n=2}^{\infty}\, e^{-\ell(n\!-\!1)}
$$
for $\,z\in(2,3)\,$, so
$$
\eta_+(z)-\eta_-(z) \,<\,{5\over\pi}\; {e^{-\ell}\over 1-e^{-\ell}}
$$
and the allowed corridor shrinks exponentially with increasing
$\,\ell\,$. On the other hand, the function
$$
g_{\theta}(u)\,:=\, {\sin u\over \cos u-\cos\theta\ell}
$$
is increasing between any two zeros of its denominator. In the
intervals, where it is positive, it is estimated by the appropriate
branch of $\,\tan\left( {u\over 2}+\pi m\right)\,$ from below; if
it is negative, we have a similar estimate from above with $\,\tan\,$
replaced by $\,-\cot\,$. Hence independently of $\,\theta\,$ we have either
$$
\xi_0(\vec a,\theta;z)\,\ge\, {\sin^2 b\over
\pi\sqrt{z\!-\!1}}\; \tan\left( {\pi\over 2}\,\left\lbrace
{\ell\over\pi} \sqrt{z\!-\!1} \right\rbrace \right)
$$
or
$$
\xi_0(\vec a,\theta;z)\,\le\, -\,{\sin^2 b\over
\pi\sqrt{z\!-\!1}}\; \cot\left( {\pi\over 2}\,\left\lbrace
{\ell\over\pi} \sqrt{z\!-\!1} \right\rbrace \right)\,,
$$
where $\,\{\cdot\}\,$ denotes the fractional part. Putting the
estimates together, we see that the oscillating part dominates, so
for sufficiently large $\,|\alpha|\,$ there are gaps having
$\,1+\,\left( \pi m\over\ell\right)^2\,$ as one endpoint provided the
latter lies in the chosen interval. In addition, $\,\tan u+\cot u\ge
2\,$, which means that the gap between the lower and the upper bound
to
$\,\xi(\vec a,\theta;z)\,$ never closes for $\,z\in(2,3)\,$ if
$$
{5\,e^{-\ell}\over 1-e^{-\ell}}\,<\, \sqrt{2}\, \sin^2 b\,.
$$
Together we infer that {\em for any} $\,\alpha\in\R\,$, the
Hamiltonian under consideration can have {\em an arbitrary finite
number of gaps} in its spectrum provided $\,\ell\,$ is chosen large
enough.

Recall that the number of gaps is a quantity strongly dependent on
the dimension of a periodic system. For one--dimensional systems it
is generically infinite \cite[Sec.XIII.16]{RS},
\cite[Sec.III.2]{AGHH}, while for higher dimensions the {\em
Bethe--Sommerfeld conjecture} states this number is finite. This is
known to be true for regular periodic potentials --- see, \eg,
\cite{S1,S2}. If the potential is replaced by a point interaction,
the one--dimensional situation suggests that the gap closing at high
energies is slower. The Bethe--Sommerfeld conjecture is known to be
valid for two-- and three--dimensional lattices if each lattice cell
contains a single point perturbation ---
\cf\cite[Secs.III.1, III.4]{AGHH}. On the other hand, an example of a
``multidimensional" periodic graph system has been presented recently
\cite{E2,EG} which exhibits an infinite number of gaps --- or none at
all. In this connection it is interesting to ask whether a
periodically perturbed strip considered here can exhibit an
infinite number of gaps. However, we are not going to attack this
problem in the present paper.

\section{Random point perturbations}

\setcounter{equation}{0}

Again, we have no intention to develop a general theory which would
modify  the results of Sec.III.5 in \cite{AGHH} to the present situation.
Our goal is more modest; we want to investigate
numerically how the scattering changes in the above studied models if
the set of point perturbations is chosen at random at an interval of
a length $\,L\,$. We will thus investigate a family of random
operators $\,H_{\omega}\equiv H(\alpha,\vec a_{\omega}, \phi)\,$,
where $\,(a_{1,\omega}, a_{2,\omega},\dots, a_{N,\omega})\,$ and $\,
(b_{1,\omega}, b_{2,\omega},\dots, b_{N,\omega})\,$ are independent
random numbers, $\,a_{j,\omega}\in (0,L)\,,\; b_{j,\omega}\in
(0,2\pi)\,$ for all $\,j=1,\dots,N\,$.

It is widely accepted that an irregular scattering on such a family
of impurities leads to fluctuations of the corresponding S--matrix,
and consequently, of the measured sample conductance. These
mesoscopic fluctuations are generic in the sense that they do not
depend on detailed properties of the sample under investigation. Our
aim here is to find the conductance fluctuations for the present
model.

The mentioned universality is usually manifested through relations to
spectral properties of random--matrix ensembles. More specifically,
the S--matrix eigenvalues of an irregular quantum scattering system
are usually supposed to conform with those of a circular ensemble of
random matrices. This correspondence has theoretically been
demonstrated by Blumel and Smilansky for the first time using
semiclassical arguments \cite{BS}. According to it, for systems with
time--reversal symmetry $\,S\,$ corresponds to the circular orthogonal
ensemble (COE), while for systems without this symmetry its
statistical properties refer to the circular unitary ensemble (CUE).
However, as we have said, the use of the COE/CUE ensembles for
description of the mesoscopic fluctuations was justified with the
help of a semiclassical approximation \cite{BS,BM}, which implicitly
requires a large number of open channels. It is therefore
interesting to investigate the situation, when only a few transverse
modes contribute to the scattering, so the mentioned argument may not
be applied.

The quantity to study is the sample conductance which is obtained (in
the two--probe setting) from the transmission amplitudes (\ref{rt_N})
by the Landauer formula (\ref{Landauer}). A standard approach to
conductance fluctuations consists of regarding the S--matrix as a
random matrix of a dimension given by the number of open channels and
averaging the corresponding conductance over the appropriate random
ensemble; this leads to \cite{BM}:
   \begin{equation} \label{averaged conductance}
\left\langle G\right\rangle\,=\,\left\lbrace\;
\begin{array}{ll}
 \frac{M}{2}\,-\, \frac{M}{4M+2}  & \qquad \mbox{\rm for COE} \\ \\
\frac{M}{2}  & \qquad \mbox{\rm for CUE,}
\end{array} \right.
   \end{equation}
where $\,\langle G\rangle$ is the mean value and $\,M\,$ is the
number of open channels in the corresponding quantum wire. For $\,M
\to\infty\,$ we get $\,\langle G\rangle_{\rm CUE}\!-\! \langle
G\rangle_{\rm COE} = {1\over 4}\,$, \ie, the averaged conductance is
enhanced by  $\,1/4\,$, if the time reversal symmetry is broken. For
the variance $\,{\rm var}(G)\,$ the averaging over the circular
ensembles leads to
   \begin{equation} \label{variance}
{\rm var}(G)\,=\,\left\lbrace\;
\begin{array}{ll}
{M(M+1)^2\over (2M+1)^2(2M+3)} & \qquad \mbox{\rm for COE}
\\ \\
{M^2\over 4(4M^2\!-1)} & \qquad \mbox{\rm for CUE.}
\end{array} \right.
   \end{equation}
This relation shows that, in the semiclassical case, the conductance
variance is practically independent of $\,M\,$ being equal to $\,1/8\,$
and $\,1/16\,$ for COE and CUE, respectively.

   \begin{figure} \label{varcond1}
   \caption{var$(G)$ for random perturbations {\em vs.} momentum}
   \end{figure}

   \begin{figure} \label{varcond2}
   \caption{Conductance for random perturbations {\em vs.} momentum}
   \end{figure}

   \begin{figure} \label{varcond3}
   \caption{Level--spacing statistics for random--impurity scattering}
   \end{figure}

The model we are considering here offers an opportunity for testing
these relations in a situation where the full solution to the quantum
scattering problem is known and the semiclassical approximation is
not required. To be specific, we will consider the dependence of the
conductance on energy of the incomimg particle in the case when the
number of impurities and their distribution inside the waveguide is
fixed; we choose 25 impurities distributed at random inside a
waveguide segment of length 25.

One expects that for a fixed number of impurities the universality
regime described above can be valid only within a certain energy
interval. For energies which are too low the localization effects
will dominate the scattering scenario and the incoming wave will be
nearly completely reflected. On the other hand for energies above a
certain threshold direct processes will dominate and the major part of
the incident wave function will be transmitted through the sample;
the region intermediate between these two regimes is expected to be
the universality ``window".

Our numerical results support the above conjectured picture. In Figure~10
we plot $\,{\rm var}(G)\,$ as a function of the incident wave vector
$\,k\,$ in comparison with the random--matrix prediction
(\ref{variance}). The universality window is localized approximately
between $\,k=3.5\,$ and $\,k=5.5\,$. For smaller $\,k\,$ the localization
effects suppress the variance whereas for $\,k\,$ above the window
the variance depends significantly on $\,k\,$ with maxima localized at
integer values of $\,k\,$.

The conductance $\,G\,$ of the same randomly perturbed waveguide is
plotted on the Figure~11. The prediction of (\ref{conductance}) is
again shown for the sake of comparison (the dashed line). It is clear
that $\,G\,$ coincides with the random--matrix prediction only inside
the universality window; for smaller $\,k\,$ it is suppressed by
localization effects while for large $\,k\,$ it is enhanced by direct
transmission.

Finally, the level--spacing statistics of the corresponding S--matrix
is shown on Figure~12. The first picture corresponds to $\,k\,$
inside the universality window, while the other one is computed for a
larger momentum value. The prediction of the random--matrix theory
(Wigner distribution) and the Poisson distribution are plotted for
the sake of comparison; a shift towards the latter for $\,k\,$ above
the window is clearly visible.

\subsection*{Acknowledgment}

At the early stage of this work which was performed several years ago
in JINR, Dubna, we benefited from discussions with H.~Holden and
N.H.~Risebro. We want also to thank for the hospitality extended to
us: P.~E. and P.~\v S. in the Institute of Mathematics, Ruhr
University, and R.~G. in the Nuclear Physics Institute, Czech Academy
of Sciences, where later parts were done. A partial support by the
grants AS No.148409 and GACR No.202--93--1314, as well as SFB 237 and
the European Union Project ERB--CiPA 3510--CT--920704/704 are also
gratefully acknowledged.

\section*{Figure captions}

   \begin{description}
   \item{\bf Figure 1\quad} The dependence of $\xi(\vec a,z)$,
where $\vec a\,=\,(0,b)$. The dotted, dash-dotted and dashed
curves represent $b\,=\,\pi/2$, $b\,=\,\pi/6$ and $b\,=\,\pi/12$,
respectively. The full line is the leading term of asymptotics,
$\xi_{as}(\vec a,z)\,=\, -\,{1\over 4\pi}\, \ln\left( -\,{z\over 4} \right)
\,+\, {1\over 2\pi}\, \Psi(1)$.
   \item{\bf Figure 2\quad} Eigenvalue plots for different values
of $b$, $\alpha$ is the coupling constant --- \cf (2.13). The dashed,
full and dash-dotted curves represent
$b\,=\,\pi/2$, $b\,=\,\pi/6$ and $b\,=\,\pi/12$, respectively.
   \item{\bf Figure 3\quad} Unnormalized eigenfunctions for
different values of $b$, $\alpha\,=\,1$. From top to bottom, they
belong to $b\,=\,\pi/2$, $b\,=\,\pi/6$ and $b\,=\,\pi/12$,
respectively.
   \item{\bf Figure 4\quad} Pole trajectory in the $\,z$--plane.
The perturbation is located in the middle of strip,
\ie $\,b\,=\,\pi/2$. The resonance is on the third sheet of
Riemann surface. The trajectory begins at the treshold $z\,=\,9$,
which corresponds to $\alpha \to +\infty$ and runs through the
lower part of the complex plane as $\alpha$ decreases to
$-\infty$. In the inset is the dependence of the residue $\rho:=
\xi'(\vec a,z)\,$ depending on the coupling constant $\,\alpha\,$.
   \item{\bf Figure 5\quad} Eigenfunctions for three point
perturbations. The ground state, $E_{0}\,=\,-8.7051$ (bottom)
and the second excited state, $E_{2}\,=\,-7.8199$ (top).
The perturbation are placed at $(0,\pi/2)$, $(1,2\pi/3)$ and
$(2,\pi/4)$; they have the same strength $\alpha\,=\,-0.15$.
   \item{\bf Figure 6\quad} An embedded eigenvalue due to
symmetry. The ground and first excited energies of a strip with two
perturbations placed symmetrically with respect to the axis of
strip at $(0,\pi/3)$ and $(0,2\pi/3)$. As $\alpha$ increases, the
first excited state reaches the continuum and tends to $4$ as
$\alpha\,\to\,\infty$.
   \item{\bf Figure 7\quad} The cascading effect for a cluster
of eigenvalues. Four perturbations placed at $(-1,2\pi/3)$,
$(0,\pi/2)$, $(1,\pi/3)$ and $(2,2\pi/5)$; the first three have
the same strength $\alpha_{1}\,=\,\alpha_{2}\,=\,\alpha_{3}\,=\,-0.2$.
If we change $\alpha_{4}$ from $+\infty$ (very weak perturbation)
to $-\infty$ (strong perturbation), there is a critical value at
which the third excited states appears below the continuum. It
becomes more bound as $\alpha_{4}$ decreases, joins the triplet
$E_{0}, E_{1}, E_{2}$ and the ground state energy $E_{0}$ parts
the cluster and drops exponentially. The order of levels is
always preserved.
   \item{\bf Figure 8\quad} The conductance plots for a pair of
point perturbations. The plots show the dependence of conductance
$G$ on energy $z$. The perturbations are at $(0,\pi/3)$,
$(0.25,2\pi/3)$, having the same strength $\alpha\,=\,0.2$ (the
upper plot) and at $(0,6\pi/11)$, $(10,3\pi/11)$,
$\alpha_{1}\,=\,-0.25$, $\alpha_{2}\,=\,-0.2$ (the lower plot).
   \item{\bf Figure 9\quad} A strip potential due to an external
field.
   \item{\bf Figure 10\quad} var$(G)$ for random perturbations
{\em vs.} momentum.
   \item{\bf Figure 11} Conductance for random perturbations {\em
vs.} momentum.
   \item{\bf Figure 12} Level--spacing statistics for random
impurity--scattering.
   \end{description}

\end{document}